\documentclass[12pt]{article}
\usepackage{titling}
\usepackage{times}
\usepackage{amsmath,amsfonts,bm,amsthm,amssymb}
\usepackage{pdflscape,float, caption, threeparttable, booktabs}
\usepackage{graphicx, multirow, comment, xcolor}

\newtheorem{theorem}{Theorem}
\newtheorem{assumption}{Assumption}

\usepackage{geometry}
\usepackage{setspace}
\usepackage[round]{natbib}
\usepackage[hyperfootnotes=false]{hyperref}
\hypersetup{
	colorlinks,
	citecolor=black,
	linkcolor=black,
	urlcolor=black}
\newcommand\blfootnote[1]{%
	\begingroup
	\renewcommand\thefootnote{}\footnote{#1}%
	\addtocounter{footnote}{-1}%
	\endgroup
}	

\begin{document}
	
	\title{Robust and Efficient Estimation of Potential Outcome Means Under
		Random Assignment}
	\author{Akanksha Negi$^{\dag}$ and Jeffrey M. Wooldridge$^{\ddag}$}
	\maketitle

    \begin{abstract}
    	\singlespacing
        We study efficiency improvements in randomized experiments for estimating a vector of potential outcome means using regression adjustment (RA) when there are more than two treatment levels. We show that linear RA which estimates separate slopes for each assignment level is never worse, asymptotically, than using the subsample averages. We also show that separate RA improves over pooled RA except in the obvious case where slope parameters in the linear projections are identical across the different assignment levels. We further characterize the class of nonlinear RA methods that preserve consistency of the potential outcome means despite arbitrary misspecification of the conditional mean functions. Finally, we apply these regression adjustment techniques to efficiently estimate the lower bound mean willingness to pay for an oil spill prevention program in California. 
       \end{abstract} 

	
	
	\blfootnote{$^\dag$Department of Econometrics and Business Statistics, Monash University. Email: \href{mailto:akanksha.negi@monash.edu}{akanksha.negi@monash.edu}, $^\ddag$Department of Economics, Michigan State University. Email: \href{mailto:wooldri1@msu.edu}{wooldri1@msu.edu}. We would like to thank the Associate Editor and three anonymous referees for their helpful comments and suggestions.}

	\newpage

    \section{Introduction}

    In the past several decades, the potential outcomes framework has become a staple of causal inference in statistics, econometrics, and related fields. Envisioning each unit in a population under different states of intervention or treatment allows one to define treatment or causal effects without reference to a model. One merely needs the potential outcome (PO) means for subpopulations.
    
    When interventions are randomized -- as in a clinical trial [\citet{hirano2001estimation}], assignment to participate in a job training program [\citet{calonico2017women}], receiving a private school voucher [\citet{angrist2006long}], or contingent valuation studies where different bid values are randomized among individuals [\citet{carson2004valuing}] -- one can simply use the subsample means for each treatment level in order to obtain unbiased and consistent estimators of the PO means. In some cases, the precision of the subsample means will be sufficient. Nevertheless, with the availability of good predictors of the outcome or response, it is natural to think that the precision can be improved, thereby shrinking confidence intervals and making conclusions about interventions more reliable.
    
    In this paper we build on \citet{negi2021revisiting}, NW (2021) hereafter, who studied the problem of estimating the average treatment effect (ATE) under random assignment with one control and one treatment group. In the context of random sampling, NW (2021) showed that performing separate linear regressions for the two groups while estimating the ATE never does worse, asymptotically, than the simple difference in means estimator or a pooled regression adjustment estimator. These findings are complementary to \citet{lin2013agnostic}, who used a finite population framework to study linear regression adjustment and arrived at essentially the same conclusions. NW (2021) also characterized the class of nonlinear regression adjustment methods that produce consistent estimators of the ATE without any additional assumptions (except regularity conditions).

    In the current paper, we allow for $G\geq 2$ treatment levels and study the problem of joint estimation of the vector of PO means. We assume that the assignment to treatment is random -- independent of both potential outcomes and observed predictors of the POs. In terms of the assignment mechanism, we assume a Bernoulli scheme whereby each unit has an equally likely probability of being assigned to a given treatment level, $g$. Importantly, other than substantive restrictions such as random assignment, $i.i.d$ sampling, and stable unit treatment value assumption, we impose only standard regularity conditions (such as finite second moments of the covariates). In other words, the regression adjustment (RA) estimators are consistent under essentially the same assumptions as the subsample means estimator with, generally, smaller asymptotic variances. Interestingly, even if the predictors are unhelpful, or the slopes in the linear projections are the same across all groups, no asymptotic efficiency is lost from using the most general RA method. 

    The extension from the binary treatment case to more than two treatment levels is nontrivial, both in terms of derivations and scope of applications. In the binary case, NW (2021) compared variances of estimators of the scalar ATE. In the current paper, we show that differences in asymptotic variance matrices of the PO estimators are positive semi-definite. Naturally, this result implies that any linear functions of the PO means are more efficiently estimated using separate RA. In addition, nonlinear functions of PO means --such as ratios -- are also more efficiently estimated. Even in the binary
    treatment case, our results significantly improve over NW (2021).\footnote{It is not enough, even in the $G=2$ case, to use NW (2021) for each of the pairwise difference in means. Knowing that each pairwise ATE is more efficiently estimated does not imply that all linear combinations are.} 
    
    We also extend the nonlinear RA results in NW (2021) to this general PO framework with $G$ treatment levels. \citet{gail1984biased}, who study nonlinear regression for the binary treatment case, have cautioned against using certain nonlinear regression models as being biased for the ATE. \citet{imbens2015causal} also leave the impression that nonlinear RA will be more efficient but at the cost of consistency, and therefore should be avoided. We show that for particular kinds of responses-- the leading cases being binary, fractional, and nonnegative -- it is possible to consistently estimate PO means using pooled and separate RA by combining results from Quasi Maximum Likelihood Estimation (QMLE) in the linear exponential family (LEF) with features of the \textit{canonical} link function.\footnote{In the generalized linear model literature, a link function $m^{-1}(\cdot)$ relates the conditional mean of the outcome to a linear predictor such that $\mathbb{E}[Y(g)|\mathbf{X}]=m(\mathbf{X}\bm{\beta})$. A canonical link is a special link function which guarantees that the model $m(\mathbf{X}\bm{\beta})$ fits the overall mean of the outcome without necessarily being correctly specified.} While \citet{gail1984biased} discussed maximum likelihood estimation in the exponential family, they do not distinguish between canonical and non-canonical link functions. 
   	As shown in NW (2021), and extended to the multiple treatment case here, one needs to use the mean function associated with the canonical link to ensure consistency without imposing additional restrictions. \citet{guo2023generalized} reach a similar conclusion in the binary treatment case studied in NW (2021), showing that generalized linear models with canonical links are examples of imputation models that are prediction unbiased for the sample average treatment effect under the design-based framework. Unlike the linear RA case, we do not have a general result that shows separate nonlinear RA is asymptotically more efficient than the subsample means estimator. However, if we add the assumption that the conditional mean functions are correctly specified, we show that doing separate RA is (weakly) more efficient than the subsample means estimator.\footnote{\citet{cohen2024no} propose a two-step calibrated Generalized Oaxaca Blinder estimator that is efficient even under misspecification.} We further establish that separate RA attains the semiparametric efficiency bound under correct conditional mean specification. 
     
    A Monte Carlo exercise substantiates our theoretical results.\footnote{We consider three different population models. The outcomes are generated to be either fractional, non-negative, or continuous with an unrestricted support. The results for fractional and non-negative outcomes can be found in section B of the online appendix.} We find that RA estimators, both linear and nonlinear, improve over subsample means in terms of standard deviations and have small biases, even with fairly small sample sizes. Not surprisingly, the magnitude of the precision gains with RA methods generally depends on how well the covariates predict the potential outcomes. Finally, we conclude with an empirical application that uses the RA methods to estimate the lower bound mean willingness to pay for an oil spill prevention program along California's coast. 
    
    In terms of contribution, this paper provides a unified framework for studying regression adjustment in experiments, allowing for multiple treatments under the infinite population (or superpopulation) setting. Besides studying linear and nonlinear RA methods and making efficiency comparisons, we also characterize the semiparametric efficiency bound for all consistent and asymptotically linear estimators of the PO means and show that separate RA attains this bound under correct conditional mean specification. Because our results compare asymptotic variance matrices for different PO mean estimators, the efficiency results derived here extend to linear and (smooth) nonlinear functions of the PO means. We also fill an important gap in the regression adjustment literature by studying pooled nonlinear methods which have not been discussed elsewhere. We show how nonlinear pooled RA is consistent whenever nonlinear separate RA is, making it an important alternative when the researcher lacks sufficient degrees of freedom for estimating separate slopes. On the practical side, the proposed RA estimators are easy to implement with common statistical software like Stata (see online appendix D for reference). 
    
    \textbf{Related Literature:} 
    This paper contributes to a vast and rich literature on regression adjustment in experiments. In the design-based framework, \citet{freedman2008regression, freeedman2008regression2} highlighted that adjusting for covariates additively in a regression is not guaranteed to produce efficiency gains in experiments compared to the simple difference in means estimator. For a binary treatment, this debate was followed up in \citet{lin2013agnostic} which showed that separate RA is weakly more efficient than both the pooled RA and simple difference in means estimators of the ATE.\footnote{More recently, \citet*{chang2021exact} propose exact bias-corrected estimators to alleviate the finite sample concerns of RA. This is different from the focus of our paper which is interested in improving efficiency with RA and assumes that sample sizes are large enough to ensure consistent estimation of the PO means. A similar argument for bias correction in high dimensional settings can be found in \citet{lei2021regression} and \citet*{chiang2023regression}.} 
    
    Among those adopting a superpopulation approach with a binary treatment, \citet{yang2001efficiency} and \citet*{leon2003semiparametric} focus on a pre-test post-test trial with no baseline covariates. Much like the current paper, specialized to the binary treatment case with ATE as the parameter of interest, \citet{yang2001efficiency} show that the interacted estimator is the most efficient among simple difference in means and pooled RA estimators. \citet{leon2003semiparametric} identify the most efficient ATE estimator using the semiparametric theory of \citet{robins1994estimation} but do not provide any efficiency comparisons between commonly employed estimators of the ATE. \citet{tsiatis2008covariate} characterize the class of all consistent and asymptotically normal (CAN) estimators of the ATE. They further show that such estimators have a common form which can be used to rank the different linear estimators in terms of efficiency. The current paper (with $G=2$) and NW (2021) arrive at the same efficiency results with respect to linear RA as \citet{tsiatis2008covariate}. In addition, the former explicitly discuss consistency with nonlinear RA (separate and pooled) involving quasi-log-likelihoods and canonical links from the generalized linear model literature. Furthermore, this paper establishes the asymptotic efficiency of separate RA relative to subsample averages when the conditional mean functions are correctly specified. \citet{pitkin2013improved} compare the asymptotic variances of simple difference in means and linear RA (pooled and separate) under essentially a similar assumption as this paper wherein the conditional mean functions are not assumed to be linear. However, \citet{pitkin2013improved} neither discuss nonlinear RA nor conditions under which it is efficient relative to simple difference in means. Similar to \citet{leon2003semiparametric}, \citet*{zhang2008improving} also study regression adjustment from a semiparametric perspective. However, their focus is on CAN estimators of pairwise treatment comparisons in a multiarmed trial. As mentioned above, the efficiency results derived in this paper extend to linear and nonlinear functions of the PO means, including pairwise ATEs. The current paper also explicitly ranks the different linear and nonlinear RA estimators, which is not explored in \citet{zhang2008improving}. We further expand our set of results to include the semiparametric efficiency bound for all CAN estimators of the PO means and establish that SRA of the QML variety attains the efficiency bound when the conditional means are correctly specified. Another paper that studies nonlinear RA for estimating the ATE is \citet{rosenblum2010simple}. The efficiency results discussed there are nested within this paper, which additionally considers pooled nonlinear methods that have not been discussed elsewhere. An added advantage of the nonlinear estimators proposed here is that they are computationally simple and easy to implement. 
    
    In the design-based setting, \citet{guo2023generalized} and \citet{cohen2024no} also discuss nonlinear RA in binary experiments. The former proposes ATE estimators which impute the missing potential outcomes using nonlinear models that are prediction unbiased. This is similar to the ``mean-fitting'' property discussed here and in NW (2021) which is crucial for consistent estimation of the PO means. This property is  guaranteed by choosing appropriate combinations of canonical link and quasi-log-likelihood functions in the LEF estimated using QMLE. \citet{cohen2024no} extend \citet{guo2023generalized}'s efficiency result to the case of mean misspecification.\footnote{They argue that a second-step calibration of the Generalized Oaxaca Blinder estimator produces an estimator of the ATE that is efficient even under misspecification of the nonlinear model.} 
   
    A recent strand also looks at multiarmed treatments in the context of factorial experiments under the design-based framework [\citet{zhao2022reconciling, zhao2022regression}, \citet{zhao2023covariate}, \citet{pashley2023causal}]. Such experiments are designed to accommodate multiple factors of interest in a single experiment and define treatment levels as all possible combinations of the factors involved. The typical concern in this literature is degree of freedom conservation while  estimating all main and interaction effects of these factors simultaneously. An exception is \citet{zhao2023covariate} which discusses linear RA for multiarmed experiments (factorial experiments being a special case). The efficiency results concerning linear RA in \citet{zhao2023covariate} have a direct analogue in the current paper under the superpoulation framework. For more complex experimental designs, see \citet{cytrynbaum2023covariate}, \citet{jiang2023regression}, \citet{roth2023efficient}, and \citet{chang2023design}.

    The rest of the paper is organized as follows. Section \ref{framework} describes the potential outcomes framework with $G$ treatment levels along with a discussion of the main assumptions. Section \ref{linear} presents the asymptotic variances of the linear RA estimators, whereas section \ref{asy_var} ranks them in terms of asymptotic efficiency.  Section \ref{nonlin} considers a class of nonlinear RA estimators that ensure consistency for the PO means without imposing additional assumptions. Section \ref{seb} characterizes the semiparametric efficiency bound for all regular estimators of the PO means and establishes local efficiency of separate RA. Section \ref{sims} constructs a simulation exercise for studying the finite sample behavior of the RA estimators. Section \ref{oilspill} applies the RA methods to study willingness to pay for an oil spill prevention program, and section \ref{conclusion} concludes.

    \section{Potential Outcomes Framework and Assumptions}\label{framework}
    We use the standard potential outcomes framework, also known as the Neyman-Rubin causal model. The goal is to estimate the population means of $G$ potential (counterfactual) outcomes, $Y(g)$, $%
    g=1,...,G$. Define $	\mu _{g}=\mathbb{E}\left[ Y(g)\right]$ for each $g=1,...,G.$

    The vector of assignment indicators is $\mathbf{W}=(W_{1},...,W_{G})$, where each $W_{g}$ is binary and $W_{1}+W_{2}+\cdots +W_{G}=1$. In other words, the groups are exhaustive and mutually exclusive. The setup applies to many situations, including the standard treatment-control setup with $G=2$, multiple treatment levels (with $g=1$ as the control
    group), and in contingent valuation studies where subjects are presented with a set of $G$ prices or bid values. 
    
     Next, let  $\mathbf{X}=(X_{1},X_{2},...,X_{K})$ be a vector of observed covariates of dimension $K$ which is assumed to be fixed in our asymptotic analysis. With respect to the assignment process, we make the following assumption.  
    \begin{assumption}[\textbf{Random Assignment}]
    	\label{assignment} Assignment is independent of the potential outcomes and observed covariates: $\mathbf{W}\perp \left[ Y(1),Y(2),...,Y(G),\mathbf{X}\right]$.
    	Further, $\rho _{g}\equiv \mathbb{P}(W_{g}=1)>0$ holds.
    \end{assumption}
    Assumption 1 puts us in the framework of experimental interventions. We also assume that each group, $g$, has a positive probability of being assigned such that $\rho _{1}+\rho _{2}+\cdots +\rho _{G}=1$.
    \begin{assumption}[\textbf{Random Sampling}]
    	\label{sampling} For a nonrandom integer $N$, $\big\{ \big[\mathbf{W}_{i},Y_{i}(1),Y_{i}(2),...,Y_{i}(G),$
    	$\mathbf{X}_{i} \big]$ $:i=1,2,...,N\big\}$ is independent and identically distributed.
    \end{assumption}

    The i.i.d assumption is not the only one we can make. For example, we could allow for a sampling-without-replacement scheme given a fixed sample size $N$. This would complicate the analysis because it generates a slight
    correlation across draws from the population. As discussed in NW (2021), Assumption \ref{sampling} is traditional in studying the asymptotic properties of estimators and is realistic as an approximation in cases where a sample is taken from a large population. It also forces us to account for the sampling error in $\mathbf{\bar{X}}$,
    as an estimator of $\bm{\mu }_{\mathbf{X}}=\mathbb{E}\left( \mathbf{X}%
    \right) $.
    
    A different approach is to assume that there is no sampling uncertainty and instead assume we observe the entire population. In the design--based approach, all uncertainty is in the assignment of the treatment indicators, $%
    \mathbf{W}=(W_{1},...,W_{G})$. Among others, \citet{freedman2008regression} and \citet{lin2013agnostic} take this approach in studying linear regression adjustment with a binary treatment. In many examples in economics (including the empirical example in our supplement), one does not observe the entire population and so it seems natural to study the random sampling framework. Our results for the random sampling case can be used to further study the efficiency issue in a framework that allows both sampling and assignment uncertainty,  as in \citet{abadie2020sampling}. Such a framework will affect inference  only in cases where the sample is a substantial fraction of the population.  If the sample size is small relative to the population size, the finite-population adjustments of the kind derived in \citet{abadie2020sampling} would have a trivial effect.
    
    Given Assumption \ref{sampling}, for each draw $i$ from the population we only observe
    \begin{equation}
    	Y_{i}=W_{i1}Y_{i}(1)+W_{i2}Y_{i}(2)+\cdots +W_{iG}Y_{i}(G),  \label{obsy}
    \end{equation}%
    and so the data are $	\left\{ \left( \mathbf{W}_{i},Y_{i},\mathbf{X}_{i}\right):i=1,2,...,N\right\}$. Definition of population quantities only requires us to use the random vector $\left( \mathbf{W},Y,\mathbf{X}\right)$, which represents the population.
    
    Along with assumptions \ref{assignment} and \ref{sampling}, we also assume that the stable unit treatment value assumption (SUTVA) holds. This implies that there are no spillovers or hidden variations of the treatments. Random assignment, $i.i.d$ sampling, and SUTVA are the only substantive restrictions used in this paper and will be maintained throughout. Subsequently, we assume that linear projections exist and that the central limit theorem holds for properly standardized sample averages of i.i.d random vectors. Therefore, we are implicitly imposing at least finite second moment assumptions on each $Y(g)$ and $X_{j}$. We do not make this explicit in what follows.

    \subsection{Completely Randomized Experiment}\label{cre}
    In this paper, we consider a randomized experiment where each unit has an equally likely probability, $\rho_g$, of being assigned to treatment level $g$. An assignment scheme that is common in practice is where a fixed number of units are assigned to each treatment level. This is known as a completely randomized experiment and creates correlation between draws. 
    Common analyses of a completely randomized experiment occur in the design-based framework where the sample coincides with the population (which is, naturally, assumed to be finite). Our framework is based on the availability of i.i.d draws, which includes the infinite superpopulation setting and also the finite population setting where sampling is done with replacement. An interesting question is whether the results obtained here apply to the setting of a completely randomized experiment, or settings where the population is finite and the assignments are i.i.d draws from a multinomial distribution. Other settings of interest include when the assignment might be determined in two (or more) stages. For example, clusters of units are defined, and then clusters are randomly assigned to be treated or control clusters. For the treated clusters, units within the cluster are randomly assigned to control or treatment-- as in \citet{abadie2023should} for the case of binary treatment without covariates. We leave these topics for future research.
    
    \section{Subsample Means and Linear Regression Adjustment}\label{linear}
    In this section we discuss the asymptotic variances of three estimators:\ the subsample means, separate regression adjustment, and pooled regression adjustment.\footnote{The asymptotic representation proofs of the three estimators can be found in section A of the online appendix.}
    
    \subsection{\textit{Subsample Means (SM)}}
	The simplest estimator of $\mu _{g}$ is the sample average within treatment group $g$: $	\bar{Y}_{g}=N_{g}^{-1}\sum_{i=1}^{N}W_{ig}Y_{i}=N_{g}^{-1}%
	\sum_{i=1}^{N}W_{ig}Y_{i}(g),$
    where $	N_{g}=\sum_{i=1}^{N}W_{ig}$ is a random variable in our setting. In expressing $\bar{Y}_{g}$ as a function of the $Y_{i}(g)$ we use $W_{ih}W_{ig}=0$ for $h\neq g$. Under random assignment and random sampling,
     {\small \begin{align*}
    	\mathbb{E}\left( \bar{Y}_{g}|W_{1g},...,W_{Ng},N_{g}>0\right) 
    	&=N_{g}^{-1}\sum_{i=1}^{N}W_{ig}\mathbb{E}\left[
    	Y_{i}(g)|W_{1g},...,W_{Ng},N_{g}>0\right]  \\
    	&=N_{g}^{-1}\sum_{i=1}^{N}W_{ig}\mathbb{E}\left[ Y_{i}(g)\right]=N_{g}^{-1}\sum_{i=1}^{N}W_{ig}\mu _{g}=\mu _{g},
    \end{align*}}%
    and so $\bar{Y}_{g}$ is unbiased conditional on observing a positive number of units in group $g$. By the law of large numbers, a consistent estimator of $\rho _{g}$ is $	\hat{\rho}_{g}=N_{g}/N$ which is the sample share of units in group $g$. Therefore, by the law of large numbers and Slutsky's Theorem, 
   {\small \begin{align*}
    	\bar{Y}_{g} &=\left( \frac{N}{N_{g}}\right)
    	N^{-1}\sum_{i=1}^{N}W_{ig}Y_{i}(g)\overset{p}{\rightarrow }\rho _{g}^{-1}%
    	\mathbb{E}\left[ W_{g}Y(g)\right]=\rho _{g}^{-1}\mathbb{E}\left( W_{g}\right) \mathbb{E}\left[ Y(g)\right]
    	=\mu _{g},
    \end{align*}}%
    and so $\bar{Y}_{g}$ is consistent for $\mu _{g}$. By the central limit theorem, $\sqrt{N}\left( \bar{Y}_{g}-\mu _{g}\right) $ is asymptotically normal. We need an asymptotic representation of $\sqrt{N}\left( \bar{Y}_{g}-\mu _{g}\right) $ that allows us to compare its asymptotic variance with those from regression adjustment estimators. To this end, write
    \begin{equation*}
    	Y(g) =\mu _{g}+V(g) \text{ and } \mathbf{\dot{X}} =\mathbf{X}-\bm{\mu }_{\mathbf{X}},
    \end{equation*}%
    where $\mathbf{\dot{X}}$ is demeaned using the population mean, $\bm{\mu }_{\mathbf{X}}$. The demeaning ensures that the intercept in the equation above can be directly interpreted as the PO mean. Now project each $V(g)$ linearly onto $\mathbf{\dot{X}}$: $V(g)=\mathbf{\dot{X}}\bm{\beta_g}+U(g)$.
    By construction, the population projection errors $U(g)$ have the properties: $	\mathbb{E}\left[ U(g)\right]  =0 \text{ and } 
    \mathbb{E}[\mathbf{\dot{X}}^{\prime }U(g)]  =\mathbf{0}\text{, 
    }$ for each $g$.
    Plugging in gives 
    \begin{equation*}
    	Y(g)=\mu _{g}+\mathbf{\dot{X}}\bm{\beta_g}+U(g)\text{, }g=1,...,G.
    \end{equation*}
    Importantly, by random assignment, $\mathbf{W}$ is independent of $[U(1),...,U(G),\mathbf{\dot{X}}]$. The observed outcome can be written as
    \begin{equation*}
    	Y=\sum_{g=1}^{G}W_{g}\left[ \mu _{g}+\mathbf{\dot{X}}\bm{\beta_g}+U(g)%
    	\right]. 
    \end{equation*}
	Our goal is to be able to make efficiency statements about both linear and nonlinear functions of the vector of means $\bm{\mu }=\left( \mu_{1},\mu _{2},...,\mu _{G}\right) ^{\prime }$, and so we stack the subsample means into the $G\times 1$ vector $\mathbf{\bar{Y}}$. For later comparison, it is helpful to remember that $\mathbf{\bar{Y}}$ is the vector of OLS coefficients in the regression (without an intercept): $Y_{i}\text{ on }W_{i1}\text{, }W_{i2}\text{, ..., }W_{iG}\text{, }%
	i=1,2,...,N.$
    The asymptotic representation for the subsample means estimator is given as 
    {\small \begin{equation*}
   	\sqrt{N}\left( \mathbf{\bar{Y}}-\bm{\mu }\right) =  N^{-1/2}\sum_{i=1}^{N}\left(\mathbf{L}_{i}+\mathbf{Q}_{i}\right)
    		+o_{p}(1)
    \end{equation*}} where  {\small \begin{equation}\label{lq}
    	\mathbf{L}_{i}\equiv 
    	\begin{pmatrix}
    		W_{i1}\mathbf{\dot{X}}_{i}\bm{\beta _{1}}/\rho _{1} \\ 
    		W_{i2}\mathbf{\dot{X}}_{i}\bm{\beta _{2}}/\rho _{2} \\ 
    		\vdots \\ 
    		W_{iG}\mathbf{\dot{X}}_{i}\bm{\beta _{G}}/\rho _{G}%
    	\end{pmatrix}
    	\text{ and  }
    	\mathbf{Q}_{i}\equiv 
    	\begin{pmatrix}
    		W_{i1}U_{i}(1)/\rho _{1} \\ 
    		W_{i2}U_{i}(2)/\rho _{2} \\ 
    		\vdots \\ 
    		W_{iG}U_{i}(G)/\rho _{G}%
    	\end{pmatrix}
	\end{equation}}

    For the vectors defined above, we know that by random assignment and the linear projection property, $\mathbb{E}\left( \mathbf{L}_{i}\right) =\mathbb{E}\left( \mathbf{Q}_{i}\right) =\mathbf{0}$, and $\mathbb{E}\left( \mathbf{L}_{i}\mathbf{Q}_{i}^{\prime }\right) =\mathbf{0}$. Also, because $W_{ig}W_{ih}=0$, $g\neq h$, the elements of $\mathbf{L}_{i}$ are pairwise uncorrelated; the same is true of the elements of $\mathbf{Q}_{i}$.
    
    \subsection{\textit{Separate Regression Adjustment (SRA)}}
    To motivate SRA, write the linear projection for each $g$ as 
    \begin{equation}\label{lp_sra}
    	Y(g) =\alpha _{g}+\mathbf{X} \bm{\beta_g}+U(g), \  \mathbb{E}\left[ U(g)\right] =0 \text{ and } \mathbb{E}\left[ \mathbf{X}^{\prime }U(g)\right] =\mathbf{0}.
    \end{equation}%
    It follows immediately that $\mu _{g}=\alpha _{g}+ \bm{\mu}_{\mathbf{X}} \bm{\beta_g}.$
    Consistent estimators of $\alpha _{g}$ and $\bm{\beta_g}$ are obtained from the regression: $Y_{i}\text{ on }1\text{, }\mathbf{X}_{i}\text{,\ if }W_{ig}=1,$
    which produces intercept and slopes $\hat{\alpha}_{g}$ and $\bm{\hat{\beta}_g}$. Therefore, the SRA estimator of PO mean, $\mu_g$ is given by $\hat{\mu}_g = \hat{\alpha}_g+\mathbf{\bar{X}}\bm{\hat{\beta}}_g$. Stacking all the SRA estimators for PO means into the vector, $\bm{\hat{\mu}}_{SRA}$, we obtain the following asymptotic representation
    {\small \begin{equation*}
    	\sqrt{N}\left(\bm{\hat{\mu}}_{SRA}-\bm{\mu}\right) = N^{-1/2}\sum_{i=1}^{N}(\mathbf{K}_i+\mathbf{Q}_i)+o_p(1)  	
    \end{equation*}} where $\mathbf{Q}_{i}$ is given in (\ref{lq}) and
	 {\small \begin{equation}\label{k}
		\mathbf{K}_{i}=%
		\begin{pmatrix}
			\mathbf{\dot{X}}_{i}\bm{\beta _{1}} \\ 
			\mathbf{\dot{X}}_{i}\bm{\beta _{2}} \\ 
			\vdots \\ 
			\mathbf{\dot{X}}_{i}\bm{\beta _{G}}%
		\end{pmatrix}.
	\end{equation}}%

    Again, for the vectors defined above, both $\mathbf{K}_{i}$ and $\mathbf{Q}_{i}$ have zero means, the latter by random assignment. Further, $\mathbb{E}\left( \mathbf{K}%
    _{i}\mathbf{Q}_{i}^{\prime }\right) =\mathbf{0}$ because $	\mathbb{E}[\mathbf{\dot{X}}_{i}^{\prime }W_{ig}U_{i}(g)] =%
    \mathbb{E}(W_{ig})\mathbb{E}[ \mathbf{\dot{X}}_{i}^{\prime }U_{i}(g)] =\mathbf{0}$.
    However, unlike the elements of $\mathbf{L}_{i}$, we must recognize that the elements of $\mathbf{K}_{i}$ are correlated except in the trivial case that all but one of the $\bm{\beta_g}$ are zero. It is important to note here that if  assignment is unconfounded given $\mathbf{X}$, SRA will be consistent provided that the conditional means are linear.
    
    \subsection{\textit{Pooled Regression Adjustment (PRA)}}
    Now consider the pooled estimator, $\bm{\hat{\mu}}_{PRA}$, which can be obtained as the vector of coefficients on $\mathbf{W}_{i}=\left(W_{i1},W_{i2},...,W_{iG}\right) $ from the regression: $Y_{i}\text{ on }\mathbf{W}_{i}\text{, }\mathbf{\ddot{X}}_{i},\text{ }
    i=1,2,...,N$
    where $\mathbf{\ddot{X}}_i= \mathbf{X}_i-\mathbf{\bar{X}}$ is the sample demeaned covariate vector. We refer to this as a pooled method because the coefficients on $\mathbf{\ \ddot{X}}_{i}$, say, $\bm{\check{\beta}}$, are assumed to be the same for all groups. Compared with subsample means, we  add the controls $\mathbf{\ \ddot{X}}_{i}$, but unlike SRA, the pooled method imposes the same coefficients across all $g$. The asymptotic representation is given as
     {\small \begin{equation*}
    	\sqrt{N}\left( \bm{\hat{\mu}}_{PRA}-\bm{\mu }\right)  = N^{-1/2}\sum_{i=1}^{N}\left( \mathbf{F}_{i}+\mathbf{K}_{i}+\mathbf{Q%
    	}_{i}\right) +o_{p}(1)
    \end{equation*}}%
    where $\mathbf{K}_{i}$ and $\mathbf{Q}_{i}$ are defined as before and  
    {\small \begin{equation} \label{f}
    	\mathbf{F}_{i}\equiv 
    	\begin{pmatrix}
    		\rho _{1}^{-1}\left( W_{i1}-\rho _{1}\right) \mathbf{\dot{X}}_{i}\bm{%
    			\delta_1} \\ 
    		\rho _{2}^{-1}\left( W_{i2}-\rho _{2}\right) \mathbf{\dot{X}}_{i}\bm{%
    			\delta_2} \\ 
    		\vdots  \\ 
    		\rho _{G}^{-1}\left( W_{iG}-\rho _{G}\right) \mathbf{\dot{X}}_{i}\bm{%
    			\delta_G}%
    	\end{pmatrix}
    \end{equation}}%
    and $\bm{\delta_g}=\bm{\beta_g}-\bm{\beta}$, where $\bm{\beta}$ is the linear projection of \ $Y$ on $\mathbf{\dot{X}}$. Again, by random assignment and the linear projection property, $	\mathbb{E}\left( \mathbf{F}_{i}\mathbf{K}_{i}^{\prime }\right) =\mathbb{E}\left(\mathbf{F}_{i}\mathbf{Q}_{i}^{\prime }\right) =\mathbf{0}$.
    
    \section{Comparing the Asymptotic Variances}\label{asy_var} 
    We now take the representations given above and use them to compare the asymptotic variances of the three estimators. It is helpful to summarize the conclusions reached in Section 3:
    {\small \begin{align}
    	\sqrt{N}\left( \bm{\hat{\mu}}_{SM}-\bm{\mu }\right) 
    	&=N^{-1/2}\sum_{i=1}^{N}\left( \mathbf{L}_{i}+\mathbf{Q}_{i}\right)
    	+o_{p}(1)  \label{ifsm} \\
    	\sqrt{N}\left( \bm{\hat{\mu}}_{SRA}-\bm{\mu }\right) 
    	&=N^{-1/2}\sum_{i=1}^{N}\left( \mathbf{K}_{i}+\mathbf{Q}_{i}\right)
    	+o_{p}(1)  \label{ifsra} \\
    	\sqrt{N}\left( \bm{\hat{\mu}}_{PRA}-\bm{\mu }\right) 
    	&=N^{-1/2}\sum_{i=1}^{N}\left( \mathbf{F}_{i}+\mathbf{K}_{i}+\mathbf{Q}%
    	_{i}\right) +o_{p}(1)  \label{ifpra}
    \end{align}}%
    where $\mathbf{L}_{i}$, $\mathbf{Q}_{i}$, $\mathbf{K}_{i}$, and $\mathbf{F}%
    _{i}$ are defined in (\ref{lq}), (\ref{k}) and (\ref{f}), respectively.
    
    \subsection{\textit{Comparing Separate RA to SM}}
    We now show that, asymptotically, $\bm{\hat{\mu}}_{SRA}$ is no worse than $\bm{\hat{\mu}}_{SM}$.
    \begin{theorem}[Efficiency of SRA relative to SM]
    	\label{smvsra} Under Assumptions \ref{assignment}, \ref{sampling}, SUTVA, and finite second moments used 
    	to obtain the asymptotic representations in \ref{ifsm} and \ref{ifsra}, 
    	\begin{equation*}
    		\mathrm{Avar}\left[ \sqrt{N}\left( \bm{\hat{\mu}}_{SM}-\bm{\mu }%
    		\right) \right] -\mathrm{Avar}\left[ \sqrt{N}\left( \bm{\hat{\mu}}_{SRA}-%
    		\bm{\mu }\right) \right] =\mathbf{\Omega _{L}}-\mathbf{\Omega _{K}}
    	\end{equation*}%
    	is PSD where $\bm{\Omega }_{\mathbf{L}}\equiv \mathbb{E}(\mathbf{L}_{i}\mathbf{L%
    	}_{i}^{\prime})$ and $\bm{\Omega }_{\mathbf{K}}\equiv \mathbb{E}(\mathbf{K}_{i}\mathbf{K%
    	}_{i}^{\prime})$.
    \end{theorem}
    The proof can be found in Appendix \ref{proofs}. To get an intuition for this efficiency gain, note that SRA uses covariate values across the entire sample for imputing the missing potential outcomes. This requires estimating the population mean of $\mathbf{X}$. Therefore, the asymptotic  variance of $\bm{\hat{\mu}}_{SRA}$ is influenced not only by the linear projection error but also $\mathbf{\bar{X}}$. In contrast, SM only averages observed outcomes for each treatment level  which means that its asymptotic variance ends up being a function of the variance of the subsample means, $\mathbf{\bar{X}}_g$, and the linear projection errors. So a comparison of the two estimators boils down to comparing $\mathbf{L}_i$ and $\mathbf{K}_i$ whose $g$-th elements are given by $W_{ig}\mathbf{\dot{X}}_i\bm{\beta_g}/\rho_g$ and $\mathbf{\dot{X}}_i\bm{\beta_g}$, respectively. To put it simply, SM is inefficient because it uses $\mathbf{\bar{X}}_g$ which is a consistent (due to random assignment) but less efficient estimator of the population mean compared to $\mathbf{\bar{X}}$, that is used by SRA. Adding orthogonal controls that reduce the error variance without inducing collinearity is relatively intuitive.  SRA does that without using a “misspecified” model in the sense that it does not impose incorrect restrictions: the slopes are the same. Imposing zero slopes (SM) and common slopes (PRA) can both be viewed as forms of misspecification. They don’t cause inconsistency, but it is not generally possible to compare efficiencies when using two “misspecified” models.
       
    The one case where there is no gain in asymptotic efficiency in using SRA is when $\bm{\beta_g}=\mathbf{0}$, $g=1,...,G$, in which case the covariates do not help predict any of the potential outcomes. Importantly, there is no gain in asymptotic efficiency in imposing $\bm{\beta_g}=\mathbf{0}$ when it is true. From an asymptotic perspective, it is harmless to separately estimate the $\bm{\beta_g}$ even when they are zero. When they are not all zero, estimating them leads to asymptotic efficiency
    gains.
    
    Theorem \ref{smvsra} also implies that any smooth nonlinear function of $\bm{\mu}$ is estimated more efficiently using $\bm{\hat{\mu}}_{SRA}$ by the delta method. For example, in estimating a percentage difference in means, we would be interested in $\mu _{2}/\mu _{1}$, and using SRA is asymptotically more efficient than using SM.

    \subsection{\textit{Separate RA versus Pooled RA}}
    The comparison between SRA and PRA is simple given the expressions in (\ref{ifsra}) and (\ref{ifpra}) because, as stated earlier, $\mathbf{F}_{i}$, $\mathbf{K}_{i}$, and $\mathbf{Q}_{i}$ are pairwise uncorrelated.
    
    \begin{theorem}[Efficiency of SRA relative to PRA]
    	\label{sravspra} Under Assumptions \ref{assignment}, \ref{sampling}, SUTVA, and finite second moments used to obtain the asymptotic representations in \ref{ifsra} and \ref{ifpra},
    	\begin{equation*}
    		\mathrm{Avar}\left[ \sqrt{N}\left( \bm{\hat{\mu}}_{PRA}-\bm{\mu }%
    		\right) \right] -\mathrm{Avar}\left[ \sqrt{N}\left( \bm{\hat{\mu}}_{SRA}-%
    		\bm{\mu }\right) \right] =\mathbf{\Omega _{F}}
    	\end{equation*}%
    	where $\bm{\Omega}_{\mathbf{F}}\equiv \mathbb{E}(\mathbf{F}_{i}\mathbf{F%
    	}_{i}^{\prime})$ is PSD.
    \end{theorem}
    The proof can be found in Appendix \ref{proofs}. It follows immediately from Theorem \ref{sravspra} that $\bm{\hat{\mu}}_{SRA}$ is never less asymptotically efficient than $\bm{\hat{\mu}}_{PRA}$. There are some special cases where the estimators achieve the same asymptotic variance, the most obvious being when the slopes in the linear projections are homogeneous: $	\bm{\beta_1}=\bm{\beta_2}=\cdots =\bm{\beta_g}$. As with comparing SRA with subsample means, there is no gain in efficiency from imposing this restriction when it is true. This is another fact that makes SRA attractive if the sample sizes within each treatment level are not small.
    
    Other situations where there is no asymptotic efficiency gain in using SRA are more subtle. In general, suppose we are interested in linear combinations $\tau =\mathbf{a}^{\prime } \bm{\mu }$ for a given $G\times
    1 $ vector $\mathbf{a}$. If $\mathbf{a}^{\prime } \mathbf{\Omega }_{\mathbf{F}}\mathbf{a}=0$, then $\mathbf{a}^{\prime } \bm{\hat{\mu}}_{PRA}$ is asymptotically as
    efficient as $\mathbf{a}^{\prime } \bm{\hat{\mu}}_{SRA}$ for estimating $%
    \tau $. Generally, the diagonal elements of $	\mathbf{\Omega }_{F}=\mathbb{E}\left( \mathbf{F}_{i}\mathbf{F}_{i}^{\prime
    }\right)$ are $	(1-\rho _{g})\bm{\delta_g}^{\prime } \mathbf{%
    \Omega }_{\mathbf{X}} \bm{\delta_g}/\rho _{g}$
    because $\mathbb{E}\left[ \left( W_{ig}-\rho _{g}\right) ^{2}\right] =\rho_{g}(1-\rho _{g})$. The off diagonal terms of $\mathbf{\Omega }_{\mathbf{F}}$ are $- \bm{\delta_g}^{\prime } \mathbf{\Omega }_{\mathbf{X}} \bm{\delta_h}$ because $\mathbb{E}\left[ \left( W_{ig}-\rho _{g}\right) \left( W_{ih}-\rho_{h}\right) \right] =-\rho _{g}\rho _{h}$. Now consider the case covered in NW (2021), where $G=2$ and $\mathbf{a}^{\prime }=\left(-1,1\right) $, so the parameter of interest is $\tau =\mu _{2}-\mu _{1}$ (the ATE). If $\rho _{1}=\rho _{2}=1/2$ then 
     {\small \begin{equation*}
    	\mathbf{\Omega }_{\mathbf{F}}=%
    	\begin{pmatrix}
    		\bm{\delta_1}^{\prime } \mathbf{\Omega }_{\mathbf{X}} \bm{\delta%
    			_1} & - \bm{\delta_1}^{\prime } \mathbf{\Omega }_{\mathbf{X}} 
    		\bm{\delta_2} \\ 
    		- \bm{\delta_2}^{\prime } \mathbf{\Omega }_{\mathbf{X}} \bm{%
    			\delta_1} & \bm{\delta_2}^{\prime } \mathbf{\Omega }_{\mathbf{X}} 
    		\bm{\delta_2}%
    	\end{pmatrix}
    \end{equation*}}%
    Now $\bm{\delta_2}=- \bm{\delta_1}$ because $\bm{\delta_1}= \bm{\beta_1}-( \bm{\beta_1}+ \bm{\beta_2})/2=(\bm{\beta_1}- \bm{\beta_2})/2=- \bm{\delta}_{2}$. Therefore,
     {\small \begin{equation*}
    	\mathbf{\Omega }_{\mathbf{F}}=%
    	\begin{pmatrix}
    		\bm{\delta_1}^{\prime } \mathbf{\Omega }_{\mathbf{X}} \bm{\delta%
    			_1} & \bm{\delta_1}^{\prime } \mathbf{\Omega }_{\mathbf{X}} \bm{%
    			\delta_1} \\ 
    		\bm{\delta_1}^{\prime } \mathbf{\Omega }_{\mathbf{X}} \bm{\delta%
    			_1} & \bm{\delta_1}^{\prime } \mathbf{\Omega }_{\mathbf{X}} \bm{%
    			\delta_1}%
    	\end{pmatrix}%
    ~and~ 
    	\begin{pmatrix}
    		-1 & 1%
    	\end{pmatrix}%
    	\begin{pmatrix}
    		\bm{\delta_1}^{\prime } \mathbf{\Omega }_{\mathbf{X}} \bm{\delta%
    			_1} & \bm{\delta_1}^{\prime } \mathbf{\Omega }_{\mathbf{X}} \bm{%
    			\delta_1} \\ 
    		\bm{\delta_1}^{\prime } \mathbf{\Omega }_{\mathbf{X}} \bm{\delta%
    			_1} & \bm{\delta_1}^{\prime } \mathbf{\Omega }_{\mathbf{X}} \bm{%
    			\delta_1}%
    	\end{pmatrix}%
    	\begin{pmatrix}
    		-1 \\ 
    		1%
    	\end{pmatrix}%
    	=0.
    \end{equation*}}
    Interestingly, the asymptotic equivalence of the separate and pooled OLS methods does not extend to the case $G\geq 3$. For any $G\geq 3$ with $\rho_{g}=1/G$ for all $g$, $	1-\rho _{g}=1-\frac{1}{G}=\frac{\left( G-1\right) }{G}$
    and so $\frac{1-\rho _{g}}{\rho _{g}}=G-1$. Note that $	\bm{\delta_g}=\bm{\beta_g}-\left( \bm{\beta_1}+\bm{\beta_2}+\cdots +\bm{\beta_g}\right) /G$ and it is less clear when a degeneracy occurs in the difference in asymptotic variances. So, for example, in our application which estimates a lower bound mean willingness to pay, SRA is generally more efficient than PRA even though the assignment probabilities are identical.
    
    \section{Nonlinear Regression Adjustment}\label{nonlin} 
    We now discuss a class of nonlinear regression adjustment methods that preserve consistency without adding additional assumptions other than the ones discussed in section \ref{framework} (and weak regularity conditions). In particular, we extend the setup in NW (2021) to more than two treatment levels.
    
    Typically, when the outcome is discrete or has limited support, nonlinear functions of $\mathbb{E}[Y(g)|\mathbf{X}]$ may be more appropriate and may provide a better approximation to the true conditional mean. However, consistency with nonlinear models is closely tied to functional form assumptions. In this section, we discuss nonlinear methods that are consistent for $\mathbb{E}[Y(g)]$ despite possible misspecification in $\mathbb{E}[Y(g)|\mathbf{X}]$. Not surprisingly, using a canonical link function in the context of QMLE in the linear exponential family plays a key role. We show that both separate and pooled nonlinear methods are consistent provided we choose the mean functions and objective functions appropriately.  Unlike in the linear case, we can only show that SRA improves over the subsample means estimator when the conditional means are correctly specified.

    \subsection{\textit{Separate Regression Adjustment}}
	 We model the conditional means, $\mathbb{E}\left[ Y(g)|\mathbf{X}\right]$, for each $g=1,2,...G$ to be $m(\alpha _{g}+\mathbf{X} \bm{\beta_g})$, where $m\left( \cdot \right)$ is a smooth function defined on $\mathbb{R}$.\footnote{The range of $m\left( \cdot \right) $ is chosen to reflect the nature of $Y(g)$. Given that the nature of $Y(g)$ does not change across treatment levels, we choose a common function $m\left( \cdot \right) $ across all $g$. Also, as usual, the vector $\mathbf{X}$ can include nonlinear functions (typically squares, interactions, and so on) of underlying covariates.} In the generalized linear model literature, $m^{-1}(\cdot)$ is known as the link function and relates the conditional mean of the outcome to a linear predictor. A canonical link is tied to a specific quasi-log-likelihood (QLL) in the LEF such that if one chooses certain combinations of QLLs and \textit{canonical} link functions, we obtain 
	 \begin{equation}
	 	\mathbb{E}\left[ Y(g)\right] =\mathbb{E}\left[ m(\alpha _{g}^{\ast }+\mathbf{%
	 		X}\bm{\beta_g}^{\ast })\right] ,  \label{mfp}
	 \end{equation}%
	 where $\alpha _{g}^{\ast }$ and $\bm{\beta_g}^{\ast }$ are the probability limits of the QMLE whether or not the conditional mean function, chosen to be $m(\cdot)$, is correctly specified. Table \ref{lef_comb} gives the pairs of mean function (or canonical links) and QLLs that ensure consistent estimation of PO means. To ensure consistency, the mean should have the index form common in the generalized linear models literature. 
	 
	 \begin{table}[htbp]
	 	\centering
	 	\resizebox{0.8\textwidth}{!}{\begin{threeparttable}
	 			\caption{Combinations of Means and QLLs to Ensure Consistency} \label{lef_comb}  
	 			\begin{tabular}{|l|l|l|}
	 				\hline
	 				\textbf{Support Restrictions} & \textbf{Mean Function} & \textbf{QLL}
	 				\\ \hline
	 				None & Linear & Gaussian (Normal) \\ \hline
	 				$Y(g)\in \left[ 0,1\right] $ (binary, fractional) & Logistic & Bernoulli \\ 
	 				\hline
	 				$Y(g)\in \left[ 0,B\right] $ (count, corners) & Logistic & Binomial \\ \hline
	 				$Y(g)\geq 0$ (count, continuous, corner ) & Exponential & Poisson \\ 
	 				\hline
	 			\end{tabular}
	 			\begin{tablenotes}[flushleft]
	 				\footnotesize
	 				\item This table enumerates combinations of quasi-log-likelihood and canonical link function specifications, both of which depend on the kind of restrictions placed on the support of the response variable.  The selection of these QLL and conditional mean function specifications will ensure consistent estimation of the PO means despite misspecification in the mean function. 
	 			\end{tablenotes}
	 	\end{threeparttable}}
	 \end{table}
	 The binomial QMLE is a good choice for counts with a known upper bound, even if it is individual-specific ($B_{i}$ need not be a positive integer for each $i$). It can also be applied to corner solution outcomes in the interval $[0,B_{i}]$ where the outcome is continuous on $(0,B_{i})$ but perhaps has mass at zero or $B_{i}$. The leading case is $B_{i}=B$. Note that we do not recommend a Tobit model in such cases because the Tobit estimates of $\mu_{g}$ are not robust to failure of the Tobit distributional assumptions or the Tobit form of the conditional mean. 
	 
	 This \textquotedblleft mean fitting property\textquotedblright\ given in \eqref{mfp} can be established by studying the population first order conditions (FOCs) with the choice of canonical link function,
	 {\small \begin{align}
	 	\mathbb{E}\left[ W_{g}\cdot (Y-m(\alpha _{g}^{\ast }+\mathbf{X}\bm{\beta_
	 		g}^{\ast }))\right] & =0  \label{foc_constant} \\
	 	\mathbb{E}\left[ W_{g}\cdot \mathbf{X}^{\prime }\cdot (Y-m(\alpha _{g}^{\ast
	 	}+\mathbf{X}\bm{\beta_g}^{\ast }))\right] & =\mathbf{0}. \label{foc_slope} 
	 \end{align}}%
    It follows from equations (\ref{obsy}) and (\ref{foc_constant}) that
    {\small \begin{equation}
    	\mathbb{E}\left[ W_{g}Y\right] =\mathbb{E}\left[ W_{g}Y(g)\right] =\mathbb{E}%
    	\left[ W_{g}m(\alpha _{g}^{\ast }+\mathbf{X}\bm{\beta_g}^{\ast })%
    	\right] .  \label{eq}
    \end{equation}}%
    Also, we can rewrite the right hand side using the law of iterated expectations as 
    {\small \begin{equation} \label{rhs} 
    	\mathbb{E}\left[ W_{g}m(\alpha _{g}^{\ast }+\mathbf{X}\bm{\beta_
    		g}^{\ast })\right]  =\mathbb{E}\left( \mathbb{E}\left[ W_{g}m(\alpha
    	_{g}^{\ast }+\mathbf{X}\bm{\beta_g}^{\ast })|\mathbf{X}\right]
    	\right)  =\rho _{g}\mathbb{E}\left[ m(\alpha _{g}^{\ast }+\mathbf{X}\bm{\beta_%
    		g}^{\ast })\right],
    \end{equation}}%
    where the second equality holds by random assignment. Using (\ref{eq}) and (\ref{rhs}) along with the fact that $\mathbb{E}\left[ W_{g}Y(g)\right] =\rho_{g}\mu _{g}$, we obtain the mean fitting result in equation (\ref{mfp}). Applying nonlinear RA, therefore, with multiple treatment levels is straightforward. One can obtain $\hat{\alpha}_{g}$ and $\bm{\hat{\beta}_g}$ by solving the sample analogues of conditions (\ref{foc_constant}) and (\ref{foc_slope}). For treatment level $g$, after obtaining $\hat{\alpha}_{g}$, $\bm{\hat{\beta}_g}$ by QMLE, the mean, $\mu _{g}$, is estimated as $\hat{\mu}_{g}=N^{-1}\sum_{i=1}^{N}m(\hat{\alpha}_{g}+\mathbf{X}_{i}\bm{%
    	\hat{\beta}_g})$,
    which includes linear RA as a special case. This estimator is consistent by a standard application of the uniform law of large numbers; see, for example, \citet{wooldridge2010econometric} (Chapter 12, Lemma 12.1). As in the linear case, the FOCs of the QMLEs using a canonical link in the LEF allow us to write the subsample averages as
    {\small \begin{equation*}
    	\bar{Y}_{g}=N_{g}^{-1}\sum_{i=1}^{N}W_{ig}m(\hat{\alpha}_{g}+\mathbf{X}_{i}%
    	\bm{\hat{\beta}_g})\text{, }g=1,...,G.
    \end{equation*}}%
    Given that $\hat{\mu}_{g}$ averages across all of the observations rather than just the subset of units at treatment level $g$, it seems that $\hat{\mu}_{g}$ should be asymptotically more efficient than $\bar{Y}_{g}$. Unfortunately, the proof used in the linear case does not generally go through in the nonlinear case. Nevertheless, when the conditional mean function is correctly specified, it is possible to show that nonlinear SRA improves over the subsample means estimator.
    
    \begin{theorem}[Efficiency of SRA relative to SM]\label{nsravsm}
    	Assume \ref{assignment}, \ref{sampling}, SUTVA, and finite second moments used to obtain the asymptotic representation for the subsample means estimator and add that the SRA estimator uses canonical link function, $m(\cdot )$, with a QMLE in the
    	linear exponential family. If for some $\alpha _{g}^{\ast }$, $\bm{\beta 
    		_g}^{\ast }$, 	$\mathbb{E}\left[ Y(g)|\mathbf{X}\right]=m(\alpha _{g}^{\ast }+\mathbf{X}\bm{\beta_g}^{\ast })\text{, }g=1,...,G$ with probability one, then %
    	\begin{equation*}
    		\mathrm{Avar}\left[ \sqrt{N}(\bm{\hat{\mu}}_{SM}-\bm{\mu })\right]-\mathrm{Avar}\left[ \sqrt{N}(\bm{\hat{\mu}}_{SRA}-\bm{\mu })\right] \text{is PSD}.
    	\end{equation*}%
    \end{theorem}
     The proof can be found in Appendix \ref{proofs}. Even when the means are not correctly specified, it seems that separate nonlinear estimation should improve over the subsample means estimator if the $\mathbf{X}$ are sufficiently predictive. We see that in the simulation results in Section \ref{sims}. Finally, as with the linear case, if assignment is only unconfounded conditional on X, nonlinear SRA is consistent under correct specification of the conditional mean, whereas SM is not.
    
    \subsection{\textit{Pooled Regression Adjustment}}
    In cases where $N$ is not especially large, one might, just as in the linear case, resort to a pooled method. Provided the mean/QLL\ combinations are chosen as in Table \ref{lef_comb}, PRA is still consistent under arbitrary misspecification of the mean function. To see why, write the mean function with common slopes $\bm{\beta}$, and without an intercept in the index, as $m(\gamma _{1}W_{1}+\gamma _{2}W_{2}+\cdots +\gamma _{G}W_{G}+\mathbf{X}\bm{\beta})$.
    Using a similar argument as in the previous section, the first-order conditions of the pooled QMLE include the $G$ equations
     {\small \begin{equation}
    	N^{-1}\sum_{i=1}^{N}W_{ig}\left[ Y_{i}-m(\check{\gamma}_{1}W_{i1}+\check{\gamma}_{2}W_{i2}+\cdots +\check{\gamma}_{G}W_{iG}+\mathbf{X}_{i}\bm{\check{\beta}})\right] =0\text{, }g=1,...,G.  \label{pqmle}
    \end{equation}}%
    Therefore, assuming no degeneracies, the probability limits of the estimators, indicated with a \textquotedblleft *\textquotedblright\ superscript, solve the population analogs of eq (\ref{pqmle}):
    {\small \begin{equation}
    	\mathbb{E}\left( W_{g}Y\right) =\mathbb{E}\left[ W_{g}Y(g)\right] =\mathbb{E}\left[ W_{g}m(\mathbf{W}\bm{\gamma }^{\ast }+\mathbf{X}\bm{\beta}^{\ast })\right] ,  \label{pqmle1}
    \end{equation}}%
    where $\mathbf{W}=(W_{1},W_{2},...,W_{G})$. By random assignment, $\mathbb{E}\left[ W_{g}Y(g) \right] =\rho _{g}\mu _{g}$ and by iterated expectations
    {\small \begin{equation}
    	\mathbb{E}\left[ W_{g}m(\mathbf{W}\bm{\gamma}^{\ast }+\mathbf{X}\bm{\beta }^{\ast })\right] =\mathbb{E}\left\{ \mathbb{E}\left[ W_{g}m(\mathbf{W}\bm{\gamma }^{\ast }+\mathbf{X}\bm{\beta}^{\ast })|\mathbf{X}\right]
    	\right\} .  \label{pqmle2}
    \end{equation}}%
    Using random assignment again, 
    {\small \begin{equation}
    	\mathbb{E}\left[ W_{g}m(\mathbf{W}\bm{\gamma}^{\ast }+\mathbf{X}\bm{\beta }^{\ast })|\mathbf{X}\right] =\mathbb{P}(W_{g}=1|\mathbf{X})m(\gamma_{g}^{\ast }+\mathbf{X}\bm{\beta}^{\ast })=\rho _{g}m(\gamma _{g}^{\ast}+\mathbf{X}\bm{\beta}^{\ast }).  \label{pqmle3}
    \end{equation}}%
    Combining equations (\ref{pqmle2}) and (\ref{pqmle3}) gives $\mathbb{E}\left[ W_{g}m(\mathbf{W}\bm{\gamma}^{\ast }+\mathbf{X}\bm{\beta }^{\ast })\right] =\rho _{g}\mathbb{E}\left[ m(\gamma _{g}^{\ast }+\mathbf{X}\bm{\beta}^{\ast })\right]$. Now, using $\rho _{g}>0$ and $\mathbb{E}[W_{g}Y(g)]=\rho _{g}\mu _{g}$, we have shown that equation (\ref{pqmle1}) gives us
    {\small \begin{equation}
    	\mu _{g}=\mathbb{E}\left[ m(\gamma _{g}^{\ast }+\mathbf{X}\bm{\beta}^{\ast })\right] ,  \label{npra}
    \end{equation}}%
    which shows that the $\mu _{g}$ are identified by the population mean of $m(\gamma _{g}^{\ast }+\mathbf{X}\bm{\beta}^{\ast })$ even when the conditional mean function is misspecified. Under weak regularity conditions, $\check{\gamma}_{g}$ is consistent for $\gamma _{g}^{\ast }$ and $\bm{\check{\beta}}$ is consistent for $\bm{\beta}^{\ast }$. Therefore, after the pooled QMLE estimation, we obtain the estimated means as $\check{\mu}_{g}=N^{-1}\sum_{i=1}^{N}m(\check{\gamma}_{g}+\mathbf{X}_{i}\bm{\check{\beta}}),$
    and these are consistent [\citet{wooldridge2010econometric} (Chapter 12, Lemma 12.1)]. As in the case of comparing nonlinear SRA to the SM, we have no general asymptotic efficiency results comparing nonlinear SRA to nonlinear PRA.
    
     \section{Semiparametric Efficiency Bound}\label{seb} 
     In this section, we generalize the efficiency result with regression adjustment in experiments for multiple treatments by providing an efficiency benchmark for estimating the vector of PO means. We characterize the efficient influence function and the semiparametric efficiency bound (SEB) for all regular and asymptotically linear estimators of $\bm{\mu}$. The SEB is useful as it provides the semiparametric analogue of the Cramer-Rao lower bound for parametric models. It gives us a standard against which to compare the asymptotic variance of any regular estimator of the potential outcome means, including the RA estimators studied in the previous sections.
     
     In our case, the most efficient estimator for $\bm{\mu}$ has the following asymptotically linear form 
     \begin{equation*}
     	\sqrt{N}\left(\bm{\hat{\mu}}-\bm{\mu}\right)= N^{-1/2}\sum_{i=1}^{N}\bm{\psi}_i+o_p(1)
     \end{equation*} such that $\mathbb{E}[\bm{\psi}_i] = \bm{0}$ and $\mathbb{E}[\bm{\psi}_i\bm{\psi}_i^\prime]$ is finite and non-singular. Then, the SEB is given by $\mathbf{V_{\ast}} = \mathbb{E}[\bm{\psi}_i\bm{\psi}_i^\prime]$ where the influence function is
 	\begin{align}\label{eif}
 		\bm{\psi}_i \equiv  \begin{pmatrix}
 			W_{i1} Y_i(1)/\rho_1- \rho_1^{-1}\left(W_{i1}-\rho_1\right)\mathbb{E}[Y_i(1)|\mathbf{X}_i] - \mu_1  \\
 			W_{i2} Y_i(2)/\rho_2- \rho_2^{-1}\left(W_{i2}-\rho_2\right) \mathbb{E}[Y_i(2)|\mathbf{X}_i] - \mu_2   \\
 			\vdots \\
 			W_{iG} Y_i(G)/\rho_G- \rho_G^{-1}\left(W_{iG}-\rho_G\right) \mathbb{E}[Y_i(G)|\mathbf{X}_i] - \mu_G
 		\end{pmatrix}.
 	\end{align}	

     This result follows from \citet{cattaneo2010efficient}, who studies efficient semiparametric estimation of treatment effects under the assumption of unconfoundedness. Given that we have the stronger assumption of randomized treatment, this essentially leads us to the same moment conditions as \citet{cattaneo2010efficient} with the exception that the propensity score here is a constant. 
	
	\begin{theorem}[Semiparametric efficiency of SRA] \label{thm:seb}
		If the conditional means of the potential outcomes, $\mathbb{E}[Y(g)|\mathbf{X}]$ for each $g$, are correctly specified then SRA is semiparametrically efficient.
	\end{theorem}
	The proof can be found in Appendix \ref{proofs}. The above result establishes that the separate regression adjustment estimator achieves the semiparametric efficiency bound if the true conditional mean of the POs is correctly specified. This implies that running separate linear regressions for different treatment level, as advocated in section \ref{asy_var}, is optimally efficient if the mean function, $m(\cdot)$, is truly linear. However, if $m(\cdot)$ is truly logistic or exponential,  then nonlinear SRA discussed in section \ref{nonlin}, is optimally efficient for estimating the PO means. 
	
    \section{Monte Carlo Simulations}\label{sims} 
    This section studies the finite sample properties of the different estimators of the PO means, namely, SM, PRA, SRA, and their nonlinear counterparts. For the simulations, we generate a population of one million observations and mimic the asymptotic setting of random sampling from an ``infinite'' population. The empirical distributions of the RA estimators are simulated for sample sizes $N\in\{500, 1000, 5000\}$ by randomly drawing the data vector $\left\{ \left(\mathbf{W}_{i},Y_{i},\mathbf{X}_{i}\right) :i=1,2,...,N\right\}$, without replacement, ten thousand times from the population. We consider three different population models, each of which uses a unique data generating process for the potential outcomes, but we report results only for the linear model. Results for fractional and non-negative outcomes can be found in online appendix B.\footnote{See Tables B.1-B.4 for results on fractional and non-negative outcomes.}
    
    To simulate multiple treatments, we consider potential outcomes, $Y(g)$, corresponding to three treatment states, $g=1, 2, 3$. Hence, $G=3$ for all population models. In each of the populations, the treatment vector $\mathbf{W}= \left(W_1, W_2, W_3\right)$ is generated with probability mass function defined by $\rho_g$ such that there is an equally likely probability of being assigned to a particular treatment group. 
    
    
    \subsection{\textit{Population Models}} 
    
    To compare the empirical distributions of the RA estimators, we consider the first population model which simulates $Y(g)$ for each $g$ to be continuous with an unrestricted range using a linear specification. We consider two covariates, $\mathbf{X} =\left(X_1, X_2\right)$, where $X_1$ is continuously distributed whereas $X_2$ is binary. The covariates are generated as follows: 
    \begin{align*}
    	X_1&= K_1+V \text{ and }
    	X_2 = 
    	\begin{cases}
    		1, \text{ if } K_2+\displaystyle \frac{V}{2}>0 \\ 
    		0, \text{ otherwise}%
    	\end{cases}%
    \end{align*}
    where $K_1, K_2 \sim N(0,2)$, and $V\sim N(0,1)$. 
    
    \paragraph{\textit{Population 1:} }For each $g$, $Y(g) 
    =\mathbf{\breve{X}}\bm{\gamma _{g}}+R(g)$
    where $\mathbf{\breve{X}}=\left( 1,X_{1},X_{2},X_{1}\cdot X_{2}\right) $ and $\bm{\gamma}_{g}=\left( \gamma _{0g},\gamma
    _{1g},\gamma _{2g},\gamma_{3g}\right) ^{\prime }$. Moreover, $R(g)\lvert X_{1},X_{2}\sim N(0,\sigma_{g}^{2})$ and $R(1),R(2),\text{ and }R(3)$ are correlated.\footnote{$R(1)\sim N(0,1),R(2)=3/\sqrt{2}\cdot \lbrack R(1)+V_{1}],R(3)=1/\sqrt{2}\cdot \lbrack R(1)+V_{2}]$ where $V_{1}$ and $V_{2}$ are distributed $N(0,1)$.} The parameter vector, $\bm{\gamma}_{g}$, is chosen to depict settings where covariates are either mildly predictive ($R^{2}(g)<0.5$) or highly predictive ($R^{2}(g)>0.5$) of the potential outcomes. For example, the parameter vectors, $\bm{\gamma}_{g}^{L}$ and $\bm{\gamma}_{g}^{H}$ are chosen such that they lead to low and high $R$-squared settings, respectively. 
    \begin{equation*}
    	\begin{split}
    		\bm{\gamma}_{1}^{L}& =\left( 0,0.2,0.1,0.1\right) ^{\prime }\text{ , }\bm{\gamma}_{2}^{L}=\left( 1,1.4,-1,-1\right) ^{\prime }\text{ , }\bm{\gamma}_{3}^{L}=\left( 2,-0.01,0.1,0.5\right) ^{\prime } \\
    		\bm{\gamma}_{1}^{H}& =\left( 0,1,0.2,-0.3\right) ^{\prime }\text{ , }\bm{\gamma}_{2}^{H}=\left( 1,2.5,0,-0.3\right) ^{\prime }\text{ , }\bm{\gamma}_{3}^{H}=\left( 2,1.1,-1,0.1\right) ^{\prime }
    	\end{split}
    \end{equation*}
    While estimating the PO means, we assume that the true population model is unknown. With linear RA methods, we simply regress the observed outcome on a constant and the covariates. For nonlinear RA, based on the nature of the outcomes, we use a particular combination of log-likelihood and canonical link function from the LEF. 
    
    \subsection{\textit{Discussion}}\label{dis}
    Tables \ref{tab:dgp11} and \ref{tab:dgp12} report the bias, standard deviation, standard error, and 95\% coverage probabilities for SM, PRA, and SRA across both high and low R-squared settings for population 1. Each table considers three different sample sizes. Note that in most cases, the bias of RA methods is comparable to SM, which is unbiased for the PO means. However, one may be willing to forego the bias in RA estimates in favor of efficiency and so we turn our attention to efficiency comparisons.
    
    Across both low and high $R^2$ configurations, we see that regression adjustment typically improves over subsample means. The magnitude of such a precision gain with PRA depends on how predictive the covariates are of the potential outcomes. We see that this gain is generally smaller at low $R^2$ than high $R^2$ values. For instance, in Table 2, the standard deviations of the PRA estimates are between 0.2\%-7.7\% smaller, across the different sample sizes, compared to the SM estimator at low $R^2$ settings, whereas these differences are of order 13.6\%-31.8\% for the high $R^2$ setting (see Table 3). The difference in precision between PRA and SRA depends on how heterogeneous are the slope coefficients in the true linear projections. While the standard deviation of the SRA estimates are between 2.3\%-6.9\% smaller than those of PRA in the low $R^2$ case, these differences are of order 5.7\%-19.4\% for the high $R^2$ case. Note that comparisons made with standard error estimates give us similar ranges, since large sample approximations are quite good. For fractional and non-negative outcomes, nonlinear RA generally seems to improve over linear RA methods and SM (see Tables B.1-B.4 in the online appendix). 
	\begin{table}[htbp]
	  \centering
	  \resizebox{0.9\textwidth}{!}{
		\begin{threeparttable}
			\caption{Monte Carlo results for RA estimators under population 1 (linear)  for a  low $R^2$} 	\label{tab:dgp11}
			\begin{tabular}{c|c|ccc|ccc|ccc}
				\toprule
				\multicolumn{11}{c}{\textbf{Low $\bm{R^2}$}} \\
			   \midrule
			   \multicolumn{1}{r|}{\multirow{2}[2]{*}{\textbf{Estimator}}} & \textbf{N}    & \multicolumn{3}{c|}{\textbf{500}} & \multicolumn{3}{c|}{\textbf{1000}} & \multicolumn{3}{c}{\textbf{5000}} \\
					\cmidrule{3-11}          &       & \textbf{$\bm{\mu_1}$} & \textbf{$\bm{\mu_2}$} & \textbf{$\bm{\mu_3}$} & \textbf{$\bm{\mu_1}$} & \textbf{$\bm{\mu_2}$} & \textbf{$\bm{\mu_3}$} & \textbf{$\bm{\mu_1}$} & \textbf{$\bm{\mu_2}$} & \textbf{$\bm{\mu_3}$} \\
			   \midrule
	   			 \multicolumn{1}{l|}{\textbf{SM}} & \textbf{Bias}  & 0.0035 & -0.0051 & -0.0031 & 0.0026 & -0.0030 & -0.0024 & 0.0021 & -0.0019 & -0.0031 \\
	          & \textbf{Sd}    & 0.0899 & 0.2948 & 0.0982 & 0.0631 & 0.2075 & 0.0692 & 0.0279 & 0.0926 & 0.0309 \\
	          & \textbf{Se}    & 0.0894 & 0.2925 & 0.0978 & 0.0633 & 0.2072 & 0.0693 & 0.0283 & 0.0927 & 0.0310 \\
	          & \textbf{Cr}    & 0.9468 & 0.9470 & 0.9461 & 0.9512 & 0.9503 & 0.9496 & 0.9536 & 0.9491 & 0.9488 \\
	          &       &       &       &       &       &       &       &       &       &  \\
	    \multicolumn{1}{l|}{\textbf{PRA}} & \textbf{Bias}  & 0.0016 & -0.0050 & -0.0013 & 0.0011 & -0.0036 & -0.0003 & 0.0006 & -0.0026 & -0.0008 \\
	          & \textbf{Sd}    & 0.0884 & 0.2730 & 0.0978 & 0.0620 & 0.1914 & 0.0686 & 0.0275 & 0.0860 & 0.0309 \\
	          & \textbf{Se}    & 0.0880 & 0.2702 & 0.0975 & 0.0622 & 0.1916 & 0.0690 & 0.0278 & 0.0858 & 0.0309 \\
	          & \textbf{Cr}    & 0.9464 & 0.9480 & 0.9490 & 0.9505 & 0.9512 & 0.9490 & 0.9529 & 0.9477 & 0.9500 \\
	          &       &       &       &       &       &       &       &       &       &  \\
	    \multicolumn{1}{l|}{\textbf{SRA}} & \textbf{Bias}  & 0.0026 & -0.0049 & -0.0022 & 0.0018 & -0.0043 & -0.0013 & 0.0013 & -0.0034 & -0.0020 \\
	          & \textbf{Sd}    & 0.0823 & 0.2661 & 0.0923 & 0.0578 & 0.1857 & 0.0646 & 0.0256 & 0.0841 & 0.0290 \\
	          & \textbf{Se}    & 0.0816 & 0.2621 & 0.0912 & 0.0579 & 0.1860 & 0.0648 & 0.0259 & 0.0834 & 0.0290 \\
	          & \textbf{Cr}    & 0.9452 & 0.9450 & 0.9454 & 0.9495 & 0.9511 & 0.9490 & 0.9535 & 0.9467 & 0.9488 \\
	    \bottomrule
	    \end{tabular}%
	  	 \begin{tablenotes}[flushleft]
		  	\footnotesize
		  	\item [a] Here SM refers to subsample means, PRA is pooled regression adjustment, and SRA is separate regression adjustment estimator.
		  	\item[b] Empirical distributions are generated with 10,000 Monte Carlo repetitions.
		  	\item[c] Row labeled `Se' stands for average of the estimated standard error based on the asymptotic variance formulas and `Cr' stands for 95\% coverage probability/rate. 
		  	\item[d] The true population mean vector is: $\bm{\mu} = (0.0588, 0.3999, 2.0969)$ 
		  	\item[e] Low $R^2$ corresponds to $(R^2_1, R^2_2, R^2_3) = \left(0.2405, 0.2865, 0.1839\right)$ 
	  	\end{tablenotes}
	\end{threeparttable}}%
	\end{table}%

	\begin{table}[htbp]
	  \centering
	  \resizebox{0.9\textwidth}{!}{
	  	\begin{threeparttable}
	  		\caption{Monte Carlo results for RA estimators under for population 1 (linear) for high $R^2$}	\label{tab:dgp12}
	  		\begin{tabular}{c|c|ccc|ccc|ccc}
	  			\toprule
	  			\multicolumn{11}{c}{\textbf{High $\bm{R^2}$}} \\
	  			\midrule
	  			\multirow{2}[4]{*}{\textbf{Estimator}} & \textbf{N} & \multicolumn{3}{c|}{\textbf{500}} & \multicolumn{3}{c|}{\textbf{1,000}} & \multicolumn{3}{c}{\textbf{5,000}} \\
	  			\cmidrule{3-11}          &       & \textbf{$\bm{\mu_1}$} & \textbf{$\bm{\mu_2}$} & \textbf{$\bm{\mu_3}$} & \textbf{$\bm{\mu_1}$} & \textbf{$\bm{\mu_2}$} & \textbf{$\bm{\mu_3}$} & \textbf{$\bm{\mu_1}$} & \textbf{$\bm{\mu_2}$} & \textbf{$\bm{\mu_3}$} \\
	    \midrule
	    \multicolumn{1}{l|}{\textbf{SM}} & \textbf{Bias}  & 0.0056 & -0.0038 & -0.0042 & 0.0044 & 0.0003 & -0.0051 & 0.0041 & 0.0011 & -0.0068 \\
	          & \textbf{Sd}    & 0.1713 & 0.4711 & 0.2151 & 0.1201 & 0.3317 & 0.1528 & 0.0531 & 0.1474 & 0.0684 \\
	          & \textbf{Se}    & 0.1692 & 0.4684 & 0.2143 & 0.1198 & 0.3314 & 0.1517 & 0.0536 & 0.1484 & 0.0678 \\
	          & \textbf{Cr}    & 0.9473 & 0.9485 & 0.9514 & 0.9483 & 0.9491 & 0.9484 & 0.9542 & 0.9529 & 0.9506 \\
	          &       &       &       &       &       &       &       &       &       &  \\
	    \multicolumn{1}{l|}{\textbf{PRA}} & \textbf{Bias}  & -0.0001 & -0.0032 & 0.0011 & -0.0003 & -0.0015 & 0.0013 & -0.0009 & -0.0010 & 0.0002 \\
	          & \textbf{Sd}    & 0.1470 & 0.3574 & 0.1482 & 0.1031 & 0.2508 & 0.1046 & 0.0459 & 0.1119 & 0.0467 \\
	          & \textbf{Se}    & 0.1460 & 0.3542 & 0.1481 & 0.1032 & 0.2509 & 0.1047 & 0.0461 & 0.1123 & 0.0468 \\
	          & \textbf{Cr}    & 0.9466 & 0.9464 & 0.9510 & 0.9492 & 0.9494 & 0.9484 & 0.9502 & 0.9495 & 0.9514 \\
	          &       &       &       &       &       &       &       &       &       &  \\
	    \multicolumn{1}{l|}{\textbf{SRA}} & \textbf{Bias}  & 0.0025 & -0.0029 & -0.0001 & 0.0016 & -0.0024 & 0.0001 & 0.0012 & -0.0021 & -0.0011 \\
	          & \textbf{Sd}    & 0.1187 & 0.3340 & 0.1388 & 0.0840 & 0.2333 & 0.0978 & 0.0370 & 0.1049 & 0.0440 \\
	          & \textbf{Se}    & 0.1180 & 0.3301 & 0.1387 & 0.0836 & 0.2340 & 0.0982 & 0.0374 & 0.1048 & 0.0440 \\
	          & \textbf{Cr}    & 0.9503 & 0.9465 & 0.9489 & 0.9486 & 0.9521 & 0.9490 & 0.9526 & 0.9484 & 0.9488 \\
	    \bottomrule
	    \end{tabular}%
		  \begin{tablenotes}[flushleft]
		  	\footnotesize
		  	\item [a] SM refers to subsample means, PRA is pooled regression adjustment, and SRA is separate regression adjustment.
		  	\item[b] Empirical distributions are generated with 10,000 Monte Carlo repetitions.
		  	\item[c] Row labeled `Se' stands for average of the estimated standard error based on the asymptotic variance formulas and `Cr' stands for 95\% coverage probability/rate. 
		  	\item[d] The true population mean vector is: $\bm{\mu} = (0.0698, 0.9654, 1.5069)$.
		  	\item[e] High $R^2$ corresponds to an $(R^2_1, R^2_2, R^2_3) = \left(0.7671, 0.7517, 0.8680\right)$. 
		  \end{tablenotes}
		\end{threeparttable}}%
	\end{table}%

\section{Application to California Oil Spill Study}\label{oilspill}
This section applies regression adjustment to estimate the lower bound mean willingness-to-pay (WTP) using survey data  from the California oil spill study that can be found in \citet{carson2004valuing}.\footnote{See online appendix C on how RA methods can be applied to estimate the lower bound on mean WTP.}  The study implements a contingent valuation survey to assess the value of damages to natural resources from future oil spills along California's central coast. 
The survey provides respondents with the choice of voting for or against a governmental program that would prevent natural resource injuries to shorelines and wildlife along the coast over the next decade. In return, the public is asked to pay a one-time lump-sum income tax surcharge for setting up the program.

The main survey sample used to elicit the yes or no votes was conducted by Westat, Inc. The data are a random sample of 1,085 interviews conducted with English speaking households where the respondent is 18 years or older and lives in private residence that is either owned or rented. Each respondent is randomly assigned one of five tax amounts: \$5, \$25, \$65, \$120, or \$220 and a choice of \textquotedblleft yes\textquotedblright\ or \textquotedblleft no\textquotedblright\ is recorded at the assigned amount.

The survey also collects data on five major characteristics of the respondents' household which are economic, demographic, preferences and attitudes towards the environment, interest in and use of the affected natural resources, evaluations of the expected harm and prevention program, and interpretations of the payment mechanism. The economic and demographic variables include log-income of the household, whether household pays state taxes in 1994, and resides in the central coast primary sampling unit. Preference towards environment include indicator variables reflecting the importance of preventing oil spills in coastal areas, spending to protect wildlife, and whether respondents consider themselves to be environmentalists or environmental activists. The attitudinal characteristics relate to respondents' attitudes towards government programs such as whether people consider spending to be important and whether respondents consider taxes to be the appropriate payment method for protecting the environment. Variables reflecting interest and use of the affected natural resources are: driving along the central coast on highway 1 and familiarity with at least one of the five species of birds most often harmed by past oil spills. Variables measuring respondents' evaluations of the expected harm and prevention program include identifying people who think oil spills over the next decade would cause more harm than mentioned in survey, who think oil spills would cause less harm, those who believe in the program's effectiveness in achieving the desired goal, and those who have concerns regarding program's effectiveness. The final set of variables relate to respondents' interpretation of the payment mechanism, such as whether respondents believe the tax to not be limited to one year and those who protested that either the oil companies should pay for the program or those who thought that the companies would pass the costs to consumers in the form of higher gas and oil prices.

Table \ref{bidsum} reports the proportion of respondents randomly assigned to the different bid amounts where we see approximately  the same number of people at each bid value.  Finally, Table \ref{te_ra} provides estimates of the PO means using linear and nonlinear RA estimators. These control for the respondent characteristics described above. We then use the PO mean estimates to obtain a lower bound on mean WTP.

In terms of the standard errors, we see a ranking among the linear RA estimators. Standard errors for the linear SRA estimator are between 0.3\%-3\% smaller than those for linear PRA estimates. Except in one case, nonlinear SRA standard errors are between 0.5\%-1.5\% smaller than for the linear SRA ones. Finally, nonlinear PRA standard errors are between 0\%-5\% smaller than those for the nonlinear SRA estimates. In this application, using pooled logistic regression produces the most efficiency gains over the usual SM estimator, also known as ABERS estimator introduced by \citet{ayer1955empirical} in contingent valuation studies--see the online appendix for reference.
\begin{table}[htbp]
	\centering
	\resizebox{0.35\columnwidth}{!}{\begin{threeparttable}
		\caption{Proportion of respondents randomly assigned to the different bid amounts}
		\begin{tabular}{ccc}
			\toprule
			\textbf{Tax} & \textbf{Total respondents} & \textbf{\%} \\
			\midrule
			\$5     & 219   & 20.2 \\
			\$25    & 216   & 19.9 \\
			\$65    & 241   & 22.2  \\
			\$120   & 181   & 16.7 \\
			\$220   & 228   & 21.0 \\
			\midrule
			\textbf{Total} & \textbf{1,085} &  \textbf{100}  \\
			\bottomrule
		\end{tabular}\label{bidsum}%
	\end{threeparttable}}
\end{table}%
\begin{table}[htbp]	
	\centering
		\resizebox{0.64\columnwidth}{!}{\begin{threeparttable}
		\caption{Lower bound mean willingness to pay estimate using ABERS and regression adjustment estimators}
		\begin{tabular}{cccccc}
			\toprule
			& \multicolumn{5}{c}{\textbf{PO means}} \\
			\cmidrule{2-6}          & \multicolumn{3}{c}{\textbf{Linear}} & \multicolumn{2}{c}{\textbf{Nonlinear}} \\
			\cmidrule{2-6}    \textbf{Bids} & \textbf{SM } & \textbf{PRA} & \textbf{SRA} & \textbf{PRA} & \textbf{SRA} \\
			\midrule
			\$5     & 0.6894 & 0.6841 & 0.6829 & 0.6841 & 0.6844 \\
			& (0.0313) & (0.0289) & (0.0284) & (0.0288) & (0.0288) \\
			\$25    & 0.5694 & 0.5874 & 0.5899 & 0.5893 & 0.5899 \\
			& (0.0338) & (0.0311) & (0.0310) & (0.0299) & (0.0308) \\
			\$65    & 0.4855 & 0.484 & 0.4815 & 0.4862 & 0.4862 \\
			& (0.0323) & (0.0299) & (0.0297) & (0.0286) & (0.0293) \\
			\$120   & 0.4033 & 0.3806 & 0.3828 & 0.3804 & 0.3813 \\
			& (0.0365) & (0.0338) & (0.0329) & (0.0328) & (0.0334) \\
			\$220   & 0.2895 & 0.2972 & 0.2921 & 0.294 & 0.2925 \\
			& (0.0301) & (0.0290) & (0.0290) & (0.0285) & (0.0286) \\
			\midrule
			\textbf{WTP } & 85.3852 & 85.1792 & 84.74 & \multicolumn{1}{r}{84.9763} & 84.8894 \\
			& (3.9051) & (3.8356) & (3.8141) & (3.6102) & (3.7947) \\
			\midrule
			\textbf{Observations} & \textbf{1,085} & \textbf{1,085} & \textbf{1,085} & \textbf{1,085} & \textbf{1,085} \\
			\bottomrule
		\end{tabular}\label{te_ra}%
		\begin{tablenotes}[flushleft]
			\footnotesize
			\item[a] Robust standard errors are in parentheses. For the WTP estimate, these are estimated using the \texttt{lincom} command in Stata (see online appendix for details). 
			\item[b]SM refers to subsample means estimator. This simply averages the `yes' responses to estimate the PO mean at a particular bid amount. PRA refers to the estimator that uses pooled regression adjustment for estimating mean bid amounts, and SRA refers to separate regression adjustment. 
			\item[c]The lower bound mean WTP estimate is calculated using  $\sum_{g=1}^{G}(b_{g}-b_{g-1})\hat{\mu} _{g}$ where ABERS estimate uses $\hat{\mu}_{g} = \bar{Y}_{g}$, and the regression adjustment estimators use $\hat{\mu}_{g} = \hat{\mu}_{PRA}$, $\hat{\mu}_{g} = \hat{\mu}_{SRA}$, $\hat{\mu}_{g} = \hat{\mu}_{\text{PRA-logit}}$, and $\hat{\mu}_{g} = \hat{\mu}_{\text{SRA-logit}}$.
		\end{tablenotes}
	\end{threeparttable}}
\end{table}%

\section{Conclusion} \label{conclusion} 
Building on the binary treatment case in NW (2021), we study efficiency improvements with RA when there are more than two treatment levels. In particular, we consider the case of random assignment of $G$ treatment levels. We show that jointly estimating the vector of potential outcome means using linear SRA, which allows for separate slopes for the different assignment levels, is asymptotically never worse than just using subsample averages; this result improves on the earlier work even when $G=2$. One case when there is no gain in asymptotic efficiency from using SRA is when the slopes are all zero. In other words, when the covariates are not predictive of the potential outcomes, then using separate slopes does not produce more precise estimates than subsample averages. We also show that SRA is generally more efficient compared to PRA, unless the slopes in true linear projections are homogeneous. In this case, using SRA to estimate the vector of PO means is harmless. 

In addition, we also extend the nonlinear RA results in NW (2021) to multiple treatment levels. We show that nonlinear RA of the QML variety is consistent if one chooses the conditional mean and objective functions appropriately from the linear exponential family of quasi-likelihoods. Furthermore, we also characterize the semiparametric efficiency bound for estimating the vector of PO means and show that SRA attains this bound, when the conditional means are correctly specified.

In simulations, we find that RA can provide substantial improvements over subsample means. Naturally, the magnitude of this precision gain generally depends on how strongly covariates predict the potential outcomes. As an illustration, we apply the different RA estimators to estimate the lower bound mean willingness to pay for a contingent valuation study which randomized different bid amounts to measure the cost of oil spills along California's coast. We find that the lower bound is estimated more efficiently when we use SRA rather than the commonly used ABERS estimator, which uses subsample averages for the PO means. 

	\singlespacing
	\bibliographystyle{ecta.bst}
	\bibliography{Bibliography}
	
	\appendix
	\numberwithin{equation}{section}
	\numberwithin{table}{section}
	
	\section{Proofs}\label{proofs}
	\begin{proof}[Theorem \ref{smvsra}]
		From (\ref{ifsm}), (\ref{ifsra}), $\mathbb{E}\left( \mathbf{L}%
		_{i}\mathbf{Q}_{i}^{\prime }\right) =\mathbf{0}$, and $\mathbb{E}\left( 
		\mathbf{K}_{i}\mathbf{Q}_{i}^{\prime }\right) =\mathbf{0}$, it follows that $\mathrm{Avar}\left[ \sqrt{N}\left( \bm{\hat{\mu}}_{SM}-\bm{\mu }\right) \right] =\bm{\Omega }_{\mathbf{L}}+\bm{\Omega }_{\mathbf{Q}}$ and $\mathrm{Avar}\left[ \sqrt{N}\left(\bm{\hat{\mu}}_{SRA}-\bm{\mu }\right) \right] =\bm{\Omega }_{\mathbf{K}}+\bm{\Omega }_{\mathbf{Q}}$ where $\bm{\Omega }_{\mathbf{L}}=\mathbb{E}\left( \mathbf{L}_{i}\mathbf{L%
		}_{i}^{\prime }\right) $ and so on. Therefore, to show that $\mathrm{Avar}%
		\left[ \sqrt{N}\left( \bm{\hat{\mu}}_{SRA}-\bm{\mu }\right) \right] $
		is smaller (in the matrix sense), we must show $\mathbf{\Omega }_{\mathbf{L}}-\mathbf{\Omega }_{\mathbf{K}}$ is PSD.
		The elements of $\mathbf{L}_{i}$ are uncorrelated because $W_{ig}W_{ih}=0$
		for $g\neq h$. The variance of the $g^{th}$ element is
		{\small \begin{equation*}
			\mathbb{E}\left[ \left( W_{ig}\mathbf{\dot{X}}_{i}\bm{\beta_g}/\rho
			_{g}\right) ^{2}\right] =\mathbb{E}\left( W_{ig}\right) \rho _{g}^{-2}\mathbb{E}\left[
			\left( \mathbf{\dot{X}}_{i}\bm{\beta_g}\right) ^{2}\right] =\rho
			_{g}^{-1}\mathbb{E}\left[ \left( \mathbf{\dot{X}}_{i}\bm{\beta%
				_g}\right) ^{2}\right] =\rho _{g}^{-1}\bm{\beta_g}^{\prime }\bm{%
				\Omega }_{\mathbf{X}}\bm{\beta_g}.
		\end{equation*}}%
		Therefore,
		{\small \begin{eqnarray*}
			\mathbb{E}\left( \mathbf{L}_{i}\mathbf{L}_{i}^{\prime }\right) =%
			\begin{pmatrix}
				\frac{\bm{\beta_1}^{\prime }\bm{\Omega }_{\mathbf{X}}\bm{\beta%
					_1}}{\rho _{1}} & 0 & \cdots & 0 \\ 
				0 & \frac{\bm{\beta_2}^{\prime }\bm{\Omega }_{\mathbf{X}}\bm{\beta%
					_2}}{\rho _{2}} & \ddots & \vdots \\ 
				\vdots & \ddots & \ddots & 0 \\ 
				0 & \cdots & 0 & \frac{\bm{\beta_G}^{\prime }\bm{\Omega }_{\mathbf{X}}%
				\bm{\beta_G}}{\rho _{G}}%
			\end{pmatrix}
			\mathbf{B}=\mathbf{B}^{\prime } \left[ \mathbf{R}
			\otimes \mathbf{\Omega }_{\mathbf{X}}\right] \mathbf{B}
		\end{eqnarray*}}%
		where
		{\small \begin{equation*}
			\mathbf{B}=%
			\begin{pmatrix}
				\bm{\beta_1} & 0 & \cdots & 0 \\ 
				0 & \bm{\beta_2} & \ddots & \vdots \\ 
				\vdots & \ddots & \ddots & 0 \\ 
				0 & \cdots & 0 & \bm{\beta_G}%
			\end{pmatrix} \text{ and } \mathbf{R} = \begin{pmatrix}
			\rho _{1}^{-1} & 0 & \cdots & 0 \\ 
			0 & \rho _{2}^{-1} & \ddots & \vdots \\ 
			\vdots & \ddots & \ddots & 0 \\ 
			0 & \cdots & 0 & \rho _{G}^{-1}%
		\end{pmatrix}
		\end{equation*}}%
		For the variance matrix of $\mathbf{K}_{i}$, $	\mathbb{V}\left( \mathbf{\dot{X}}_{i}\bm{\beta_g}\right) =\bm{%
			\beta_g}^{\prime }\bm{\Omega }_{\mathbf{X}}\bm{\beta_g}$ and $\mathbb{C}(\mathbf{\dot{X}}_{i}\bm{\beta }_{g},\mathbf{\dot{X}}_{i}%
		\bm{\beta_h})=\bm{\beta_g}^{\prime }\bm{\Omega }_{%
		\mathbf{X}}\bm{\beta_h}$.
		Therefore, one may write
		{\small \begin{equation*}
			\mathbb{E}\left( \mathbf{K}_{i}\mathbf{K}_{i}^{\prime }\right) 
			=\mathbf{B}^{\prime }\left[ \left( \mathbf{j}_{G}\mathbf{j}%
			_{G}^{\prime }\right) \otimes \mathbf{\Omega }_{\mathbf{X}}\right] \mathbf{B}
		\end{equation*}}%
		where $\mathbf{j}_{G}^{\prime }=(1,1,...,1)$. Therefore, the comparison we
		need to make is $ \mathbf{R}
		\otimes \mathbf{\Omega }_{\mathbf{X}}\text{ versus }\left( \mathbf{j}_{G}%
		\mathbf{j}_{G}^{\prime }\right) \otimes \mathbf{\Omega }_{\mathbf{X}}$. 
		That is, we need to show $	\left[ 
		\mathbf{R}
		-\left( \mathbf{j}_{G}\mathbf{j}_{G}^{\prime }\right) \right] \otimes 
		\mathbf{\Omega }_{\mathbf{X}}$
		is PSD. The Kronecker product of two PSD matrices is also PSD, so it
		suffices to show $ \mathbf{R}
		-\left( \mathbf{j}_{G}\mathbf{j}_{G}^{\prime }\right)$
		is PSD when the $\rho _{g}$ add to unity. Let $\mathbf{a}$ be any $G\times 1$
		vector. Then $	\mathbf{a}^{\prime }%
		\mathbf{R}
		\mathbf{a}=\sum_{g=1}^{G}a_{g}^{2}/\rho _{g} \implies 	\mathbf{a}^{\prime }\left( \mathbf{j}_{G}\mathbf{j}_{G}^{\prime }\right) 
		\mathbf{a}=\left( \mathbf{a}^{\prime }\mathbf{j}_{G}\right) ^{2}=\left(
		\sum_{g=1}^{G}a_{g}\right) ^{2}$. So we have to show that $\sum_{g=1}^{G}a_{g}^{2}/\rho _{g}\geq \left( \sum_{g=1}^{G}a_{g}\right) ^{2}$.
		Define vectors $\mathbf{b}=\left( a_{1}/\sqrt{\rho _{1}},a_{2}/\sqrt{\rho_{2}},...,a_{G}/\sqrt{\rho _{G}}\right) ^{\prime }$ and $\mathbf{c}=\left(\sqrt{\rho _{1}},\sqrt{\rho _{2}},...,\sqrt{\rho _{G}}\right) ^{\prime }$ and apply the Cauchy-Schwarz inequality:
		{\small \begin{equation*}
			\left( \sum_{g=1}^{G}a_{g}\right) ^{2}=\left( \mathbf{b}^{\prime }\mathbf{%
				c}\right) ^{2}\leq \left( \mathbf{b}^{\prime }\mathbf{b}\right) \left( 
			\mathbf{c}^{\prime }\mathbf{c}\right) =\left( \sum_{g=1}^{G}a_{g}^{2}/\rho
			_{g}\right) \left( \sum_{g=1}^{G}\rho _{g}\right)
			=\left( \sum_{g=1}^{G}a_{g}^{2}/\rho _{g}\right)
		\end{equation*}}%
		because $\sum_{g=1}^{G}\rho _{g}=1$. 
	\end{proof}
	
	\begin{proof}[Theorem \ref{sravspra}]	
		From (\ref{ifsra}), (\ref{ifpra}), $\mathbb{E}(\mathbf{F}_i\mathbf{K}_i^\prime)= \mathbb{E}(\mathbf{K}_i\mathbf{Q}_i^\prime) = \mathbb{E}(\mathbf{F}_i\mathbf{Q}_i^\prime)=\mathbf{0}$, it follows that $\mathrm{Avar}\left[\sqrt{N}\left(\bm{\hat{\mu}}_{SRA}-\bm{\mu }\right)\right] =\bm{\Omega }_{\mathbf{K}}+\bm{\Omega }_{\mathbf{Q}}$
		and $	\mathrm{Avar}\left[ \sqrt{N}\left( \bm{\hat{\mu}}_{PRA}-\bm{\mu }%
		\right) \right] =\bm{\Omega }_{\mathbf{F}}+\bm{\Omega }_{\mathbf{K}}+\bm{\Omega }_{\mathbf{Q}}$
		where $\bm{\Omega }_{\mathbf{K}}=\mathbb{E}\left( \mathbf{K}_{i}\mathbf{K%
		}_{i}^{\prime }\right) $ and so on. Therefore, comparing the two variance expressions, we obtain the result.
	\end{proof}

	\begin{proof}[Theorem \ref{nsravsm}]
		Let $m(\alpha_g+\mathbf{X}\bm{\beta_g})$ be the mean function associated with the canonical link function from a linear exponential family for $g=1,2,\ldots,G$.  Then the QMLE estimators, $\hat{\alpha}_g$ and $\bm{\hat{\beta}_g}$ satisfy the  first order conditions $\sum_{i=1}^{N} W_{ig}\cdot [Y_i-m(\hat{\alpha}_g+\mathbf{X}_i\bm{\hat{\beta}_g})]= 0$ and $\sum_{i=1}^{N} W_{ig}\cdot\mathbf{X}_i^\prime \cdot [Y_i-m(\hat{\alpha}_g+\mathbf{X}_i\bm{\hat{\beta}_g})] = \mathbf{0}$.
		Then, the SRA estimators of $\mu_g$ are given as $\hat{\mu}_g = \frac{1}{N}\sum_{i=1}^{N} m(\hat{\alpha}_g+\mathbf{X}_i\bm{\hat{\beta}_g})$.  For notational simplicity, let $\mathbf{\breve{X}} \equiv \left(1, \mathbf{X}\right)$, and $\bm{\delta}_g \equiv \left(\alpha_g, \bm{\beta_g}\right)^\prime$. 
		Now, 
		{\small\begin{equation*}
			\begin{split}
				\sqrt{N}(\hat{\mu}_g-\mu_g) 
				&= \frac{1}{\sqrt{N}}\sum_{i=1}^{N}\dot{m}_{ig} + \mathbf{M}^\ast_g  \sqrt{N}(\bm{\hat{\delta}}_g-\bm{\delta}^\ast_g)+o_p(1)
			\end{split}
		\end{equation*}}
		where $m_{ig}^\ast = m(\mathbf{\breve{X}}_i\bm{\delta}^\ast_g)$, $\dot{m}_{ig} = m_{ig}^\ast-\mu_g$, and $\mathbf{M}^\ast_g = \mathbb{E}\left[\nabla_{\delta_g} m(\mathbf{\breve{X}}_i\bm{\delta}^\ast_g)\right] $.
		Because of estimation via QMLE using linear exponential family with canonical mean $m(\cdot)$, $\bm{\hat{\delta}}_g$ will satisfy FOC: $	\frac{1}{N}\sum_{i=1}^{N}W_{ig}\mathbf{\breve{X}}_i^\prime\left[Y_i(g)-m(\mathbf{\breve{X}}_i\bm{\hat{\delta}}_g)\right] = \mathbf{0}$
		which implies that 
		{\small \begin{equation*}
			\sqrt{N}\left(\bm{\hat{\delta}}_g-\bm{\delta}^\ast_g\right) = \Big(\mathbb{E}\left[W_{ig}\mathbf{\breve{X}}^\prime_i\nabla_{\delta_g}m(\mathbf{\breve{X}}_i\bm{\delta}^\ast_g)\right]\Big)^{-1} N^{-1/2}\sum_{i=1}^{N}W_{ig}\mathbf{\breve{X}}^\prime_iU_i(g)+o_p(1)
		\end{equation*}}
		where $U_i(g) \equiv Y_i(g)-m(\mathbf{\breve{X}}_i\bm{\delta}_g^\ast)$. Because of random assignment, we know that $W_{ig}$ is independent of $\mathbf{\breve{X}}_i$ and therefore we can write $	\mathbb{E}\left[W_{ig}\mathbf{\breve{X}}^\prime_i\nabla_{\delta_g}m(\mathbf{\breve{X}}_i\bm{\delta}^\ast_g)\right] = \rho_g\cdot \mathbf{C}_g^\ast$ where $\rho_g \equiv P(W_{ig}=1)$ and $	\mathbf{C}_g^\ast \equiv \mathbb{E}\left[\mathbf{\breve{X}}^\prime_i\nabla_{\delta_g}m(\mathbf{\breve{X}}_i\bm{\delta}^\ast_g)\right] $. 	
		Substituting the influence function (IF) for $\bm{\hat{\delta}}_g$ into the IF for $\hat{\mu}_g$,
		{\small \begin{equation*}
			\sqrt{N}(\hat{\mu}_g-\mu_g)= N^{-1/2}\sum_{i=1}^{N}\dot{m}_{ig} + \rho_g^{-1}\mathbf{M}^\ast_g \mathbf{C}^{\ast -1}_g\Bigg[N^{-1/2}\sum_{i=1}^{N}W_{ig}\mathbf{\breve{X}}^\prime_iU_i(g)\Bigg] +o_p(1) 
		\end{equation*}}
		Hence, stacking all such IFs for $g=1,2,\ldots, G$, we obtain
		{\small \begin{equation} \label{sra.if}
			\sqrt{N}\left(\bm{\hat{\mu}}_{SRA}-\bm{\mu}\right) 
			= N^{-1/2}\sum_{i=1}^{N}\left(\mathbf{H}_i+\mathbf{J}_i\right)+o_p(1)
		\end{equation}}
		where 
		{\small \begin{equation*}
			\mathbf{H}_i = \begin{pmatrix}
				\dot{m}_{i1} \\
				\dot{m}_{i2} \\
				\vdots \\
				\dot{m}_{iG} 
			\end{pmatrix}\text{ and } \mathbf{J}_i = \begin{pmatrix}
				\rho_1^{-1}\mathbf{M}^\ast_1\mathbf{C}^{\ast-1}_1W_{i1}\mathbf{\breve{X}}^\prime_iU_i(1) \\
				\rho_2^{-1}\mathbf{M}^\ast_2\mathbf{C}^{\ast-1}_2W_{i2}\mathbf{\breve{X}}^\prime_iU_i(2)  \\
				\vdots \\
				\rho_G^{-1}\mathbf{M}^\ast_G\mathbf{C}^{\ast-1}_GW_{iG}\mathbf{\breve{X}}^\prime_iU_i(G) 
			\end{pmatrix}
		\end{equation*}}
		Now, $\sqrt{N}\left(\bar{Y}_g -\mu_g\right) =N^{-1/2} \sum_{i=1}^{N}\frac{W_{ig}}{\rho_g}\text{ }\dot{m}_{ig}+ \rho_g^{-1}\mathbf{M}^\ast_g\mathbf{C}^{\ast-1}_g N^{-1/2}\sum_{i=1}^{N}W_{ig}\mathbf{\breve{X}}_i^\prime U_i(g)+o_p(1) $.
		Similarly, stacking all such IFs for all $g=1,2,\ldots,G$, we get	
		{\small \begin{equation} \label{sm.if}
			\sqrt{N}\left(\mathbf{\bar{Y}}-\bm{\mu}\right) 
			 = N^{-1/2}\sum_{i=1}^{N}\left(\mathbf{P}_i+\mathbf{J}_i\right)+o_p(1)
		\end{equation}}
		where $\mathbf{P}_i = \begin{pmatrix}
			\frac{W_{i1}}{\rho_1}\text{ } \dot{m}_{i1}& 	\frac{W_{i2}}{\rho_2}\text{ } \dot{m}_{i2} & \ldots & 
			\frac{W_{iG}}{\rho_G}\text{ } \dot{m}_{iG}
		\end{pmatrix}^\prime$. 
	Consider equations (\ref{sra.if}) and (\ref{sm.if}) in order to compare the two estimators. We see that both equations have the same vector component, $\mathbf{J}_i$. Note that if the conditional mean function is correctly specified, then, by random assignment, $	\mathbb{E}\left[U_{i}(g)|\mathbf{\breve{X}}_i, \mathbf{W}_i\right] = 0$.
		This implies, $\mathbb{E}\left(\mathbf{J}_i\mathbf{H}_i^\prime\right)=\mathbb{E}\left(\mathbf{J}_i\mathbf{P}_i^\prime\right) =\mathbf{0}$ where,	
		{\footnotesize \begin{equation*}
				\begin{split}
				\mathbb{E}\left(\mathbf{J}_i\mathbf{P}_i^\prime\right) 
					= \begin{pmatrix}
						\rho_1^{-1}\mathbf{M}_1^\ast \mathbf{C}_1^{\ast-1}\mathbb{E}\left[\mathbf{\breve{X}}_i^\prime U_i(1)\dot{m}_{i1}\right] & 0 & \ldots & 0  \\
						0& \rho_2^{-1}\mathbf{M}_2^\ast \mathbf{C}_2^{\ast-1}\mathbb{E}\left[\mathbf{\breve{X}}_i^\prime U_i(2)\dot{m}_{i2}\right] & \ldots & 0  \\
						\vdots & \vdots & \ddots &  \vdots \\
						0 & 0 & \ldots & \rho_G^{-1}\mathbf{M}_G^\ast \mathbf{C}_G^{\ast-1}\mathbb{E}\left[\mathbf{\breve{X}}_i^\prime U_i(G)\dot{m}_{iG}\right] 
					\end{pmatrix}
				\end{split} 
		\end{equation*}}
		and,
		{\footnotesize \begin{equation*}
				\begin{split}
					\mathbb{E}\left(\mathbf{J}_i\mathbf{H}_i^\prime\right) 
					= \begin{pmatrix}
						\mathbf{M}_1^\ast \mathbf{C}_1^{\ast-1}\mathbb{E}\left[\mathbf{\breve{X}}_i^\prime U_i(1)\dot{m}_{i1}\right] & \mathbf{M}_1^\ast \mathbf{C}_1^{\ast-1}\mathbb{E}\left[\mathbf{\breve{X}}_i^\prime U_i(1)\dot{m}_{i2}\right] & \ldots & \mathbf{M}_1^\ast \mathbf{C}_1^{\ast-1}\mathbb{E}\left[\mathbf{\breve{X}}_i^\prime U_i(1)\dot{m}_{iG}\right]  \\
						\mathbf{M}_2^\ast \mathbf{C}_2^{\ast-1}\mathbb{E}\left[\mathbf{\breve{X}}_i^\prime U_i(2)\dot{m}_{i1}\right] & \mathbf{M}_2^\ast \mathbf{C}_2^{\ast-1}\mathbb{E}\left[\mathbf{\breve{X}}_i^\prime U_i(2)\dot{m}_{i2}\right] & \ldots & \mathbf{M}_2^\ast \mathbf{C}_2^{\ast-1}\mathbb{E}\left[\mathbf{\breve{X}}_i^\prime U_i(2)\dot{m}_{iG}\right]  \\
						\vdots & \vdots & \ddots &  \vdots \\
						\mathbf{M}_G^\ast\mathbf{C}_G^{\ast-1}\mathbb{E}\left[\mathbf{\breve{X}}_i^\prime U_i(G)\dot{m}_{i1}\right]  & \mathbf{M}_G^\ast\mathbf{C}_G^{\ast-1}\mathbb{E}\left[\mathbf{\breve{X}}_i^\prime U_i(G)\dot{m}_{i2}\right]  & \ldots & \mathbf{M}_G^\ast \mathbf{C}_G^{\ast-1}\mathbb{E}\left[\mathbf{\breve{X}}_i^\prime U_i(G)\dot{m}_{iG}\right] 
					\end{pmatrix} 
				\end{split}
		\end{equation*}}	
		Therefore, $	\text{Avar}\big[\sqrt{N}\left(\mathbf{\bar{Y}}-\bm{\mu}\right)\big]-\text{Avar}\big[\sqrt{N}\left(\bm{\hat{\mu}}_{SRA}-\bm{\mu}\right)\big]= \mathbb{E}\left(\mathbf{P}_i\mathbf{P}_i^\prime\right)-\mathbb{E}\left(\mathbf{H}_i\mathbf{H}_i^\prime\right)$.
		Now, we can express
		{\small \begin{equation*}
			\begin{split}
				\mathbb{E}\left(\mathbf{P}_i\mathbf{P}_i^\prime\right) 
				& = \mathbb{E}\left[\begin{pmatrix}
					\dot{m}_{i1} & 0 & \ldots & 0 \\
					0 & \dot{m}_{i2} & \ldots & 0 \\
					\vdots & \vdots & \ddots &  \vdots \\
					0 & 0 & \ldots & \dot{m}_{iG}
				\end{pmatrix}\mathbf{R} \begin{pmatrix}
					\dot{m}_{i1} & 0 & \ldots & 0 \\
					0 & \dot{m}_{i2} & \ldots & 0 \\
					\vdots & \vdots & \ddots &  \vdots \\
					0 & 0 & \ldots & \dot{m}_{iG}
				\end{pmatrix}\right]\\
			\end{split}
		\end{equation*}}
		The off-diagonal terms are all zero since, $W_{ig}\cdot W_{ih} = 0$ for any $g\neq h$. Similarly, 
		{\small \begin{equation*}
			\begin{split}
				\mathbb{E}\left(\mathbf{H}_i\mathbf{H}_i^\prime\right) 
				& =\mathbb{E}\left[\begin{pmatrix}
				\dot{m}_{i1} & 0 & \ldots & 0 \\
				0 & \dot{m}_{i2} & \ldots & 0 \\
				\vdots & \vdots & \ddots &  \vdots \\
				0 & 0 & \ldots & \dot{m}_{iG}
			\end{pmatrix} \mathbf{j}_G\mathbf{j}_G^\prime\begin{pmatrix}
			\dot{m}_{i1} & 0 & \ldots & 0 \\
			0 & \dot{m}_{i2} & \ldots & 0 \\
			\vdots & \vdots & \ddots &  \vdots \\
			0 & 0 & \ldots & \dot{m}_{iG}
		\end{pmatrix} \right]
				\end{split}
		\end{equation*}}
		where $\mathbf{j}_{G}^{\prime }=(1,1,...,1)$. Therefore, $\text{Avar}\big[\sqrt{N}\left(\mathbf{\bar{Y}}-\bm{\mu}\right)\big]-\text{Avar}\big[\sqrt{N}\left(\bm{\hat{\mu}}_{SRA}-\bm{\mu}\right)\big]=$
		
		{\small \begin{equation*}
			\begin{split}	
				&= \mathbb{E}\left[\begin{pmatrix}
					\dot{m}_{i1} & 0 & \ldots & 0 \\
					0 & \dot{m}_{i2} & \ldots & 0 \\
					\vdots & \vdots & \ddots &  \vdots \\
					0 & 0 & \ldots & \dot{m}_{iG}
				\end{pmatrix}\underbrace{\left\{ 
			\mathbf{R}%
				-\left( \mathbf{j}_{G}\mathbf{j}_{G}^{\prime }\right) \right\}}_{\text{\normalfont(A.3)}}  \begin{pmatrix}
					\dot{m}_{i1} & 0 & \ldots & 0 \\
					0 & \dot{m}_{i2} & \ldots & 0 \\
					\vdots & \vdots & \ddots &  \vdots \\
					0 & 0 & \ldots & \dot{m}_{iG}
				\end{pmatrix}\right] 
			\end{split}
		\end{equation*}}
		Here we can use the result that if $\mathbf{T}$ is a real symmetric matrix that is PSD, then $\mathbf{S}^\prime \mathbf{T} \mathbf{S}$ is also real, symmetric and PSD for some real matrix $\mathbf{S}$. Moreover, expectation of a random PSD matrix is also PSD. To utilize these results here, all we need to show is that (A.3) is PSD which we have already established in the proof of Theorem \ref{smvsra}. Hence, we obtain our result. 
		\end{proof}

	\begin{proof}[Theorem \ref{thm:seb}]
		To show that linear SRA achieves the bound, let us assume that for each $g$, $Y(g) = \mu_g +\mathbf{\dot{X}}\bm{\beta}_g+U(g), \ \mathbb{E}[U(g)|\mathbf{X}] = 0.$
		 Note that one can rewrite
		{\small \begin{align*}
			\psi_{g}  = \frac{W_g}{\rho_g}\cdot \left(Y(g)-\mathbb{E}[Y(g)|\mathbf{X}]\right)+ \mathbb{E}[Y(g)|\mathbf{X}]- \mu_g  = \frac{W_g}{\rho_g}\cdot U(g)+\mathbf{\dot{X}}\bm{\beta}_g
		\end{align*}}
		Then, looking at the asymptotic representation of the linear SRA estimator, we see that the influence function\footnote{See Lemma 2 of the online appendix for the asymptotic representation of the SRA estimator.} of $\bm{\hat{\mu}}_{SRA}$ is given by $\bm{\psi}_i$ such that for each $g$, $\psi_{ig} = W_{ig} U_i(g)/\rho_g+\mathbf{\dot{X}}_i\bm{\beta}_g$ and $\textup{Avar}(\bm{\hat{\mu}}_{SRA})=\mathbf{V}_{\ast}/N$.
		
		To show that the nonlinear SRA estimator achieves the SEB, let  $\mathbb{E}[Y(g)|\mathbf{X}] = m(\alpha_g^\ast+\mathbf{X}\bm{\beta}_g^\ast) = m(\mathbf{\breve{X}}\bm{\delta}_g^\ast)$. Then, the asymptotic expansion of the nonlinear SRA estimator is $	\sqrt{N}(\bm{\hat{\mu}}_{SRA} - \bm{\mu}) = N^{-1/2}\sum_{i=1}^{N}(\mathbf{H}_i+\mathbf{J}_i)+o_p(1)$ where the $g$-th element of the influence function, $\mathbf{H}_i+\mathbf{J}_i$, is given by $	\dot{m}_{ig}+\rho_g^{-1}\mathbf{M}_g^\ast\mathbf{C}_g^{\ast-1} W_{ig}\mathbf{\breve{X}}_iU_i(g)$ where $\dot{m}_{ig} = m_{ig}^\ast-\mu_g$, $m_{ig}^\ast = m(\mathbf{\breve{X}}_i\bm{\delta}^\ast_g)$, $\mathbf{M}_g^\ast = \mathbb{E}\left[\nabla_{\delta_g} m(\mathbf{\breve{X}}_i\bm{\delta}^\ast_g)\right]$, and $	\mathbf{C}_g^\ast \equiv \mathbb{E}\left[\mathbf{\breve{X}}^\prime_i\nabla_{\delta_g}m(\mathbf{\breve{X}}_i\bm{\delta}^\ast_g)\right]$.  Now, let's define, $ f_{ig}^\ast = \partial m(\mathbf{\breve{X}}_i\bm{\delta}_g)/\partial (\mathbf{\breve{X}}\bm{\delta}_g)|_{\bm{\delta}_g = \bm{\delta}_g^\ast}$. Then, 
		{\small \begin{align*}
			\mathbf{M}_g^\ast = \begin{pmatrix}
				\mathbb{E}(f_{ig}^\ast) & \mathbb{E}(\mathbf{X}_if_{ig}^\ast)
			\end{pmatrix}  \text{ and }  	\mathbf{C}_g^\ast = \begin{pmatrix}
				\mathbb{E}(f_{ig}^\ast) & \mathbb{E}(\mathbf{X}_if_{ig}^\ast) \\
				\mathbb{E}(\mathbf{X}_i^\prime f_{ig}^\ast) & \mathbb{E}(\mathbf{X}_i^\prime \mathbf{X}_i f_{ig}^\ast)
			\end{pmatrix}
		\end{align*}}
		Using inverse rules for partitioned matrices, we know that $ \mathbf{C}_g^{\ast-1} = \begin{pmatrix}
			c_{11} & c_{12} \\
			\mathbf{c}_{21} & \mathbf{c}_{22}
		\end{pmatrix}$ where 
		{\small \begin{align*}
			c_{11} & = \mathbb{E}(f_{ig}^\ast)^{-1}\bigg\{1+\mathbb{E}(\mathbf{X}_if_{ig}^\ast)\mathbf{Q}^{-1}\mathbb{E}(\mathbf{X}_i^\prime f_{ig}^\ast)\mathbb{E}(f_{ig}^\ast)^{-1}\bigg\} & c_{12} &= -\mathbb{E}(f_{ig}^\ast)^{-1}\mathbb{E}(\mathbf{X}_if_{ig}^\ast)\mathbf{Q}^{-1} \\
			\mathbf{c}_{21}& = - \mathbf{Q}^{\prime-1}\mathbb{E}(\mathbf{X}_i^\prime f_{ig}^\ast)\mathbb{E}(f_{ig}^\ast)^{-1} & 	\mathbf{c}_{22} &=\mathbf{Q}^{-1} 	
		\end{align*}} and $\mathbf{Q} = \mathbb{E}(\mathbf{X}_i^\prime\mathbf{X}_if_{ig}^\ast)-\mathbb{E}(\mathbf{X}_i^\prime f_{ig}^\ast)\mathbb{E}(f_{ig}^\ast)^{-1}\mathbb{E}(\mathbf{X}_if_{ig}^\ast)$. Then, 
	\begin{equation*}
		\mathbf{M}_g^\ast\mathbf{C}_g^{\ast-1}= \begin{pmatrix}
			\mathbb{E}(f_{ig}^\ast)c_{11}+\mathbb{E}(\mathbf{X}_if_{ig}^\ast)\mathbf{c}_{21} & \mathbb{E}(f_{ig}^\ast)c_{12}+\mathbb{E}(\mathbf{X}_if_{ig}^\ast)\mathbf{c}_{22}
		\end{pmatrix} = \begin{pmatrix}
			1, \mathbf{0}_{1\times K}
		\end{pmatrix}. 
	\end{equation*}
This implies that $\mathbf{M}_g^\ast\mathbf{C}_g^{\ast-1} W_{ig}\mathbf{\breve{X}}_iU_i(g) = W_{ig}U_{i}(g)$  so $\dot{m}_{ig}+\rho_g^{-1}\mathbf{M}_g^\ast\mathbf{C}_g^{\ast-1} W_{ig}\mathbf{\breve{X}}_iU_i(g) = \psi_{ig}$.
	\end{proof}

\end{document}